\documentclass[aps,twocolumn,secnumarabic,nobalancelastpage,amsmath,amssymb,nofootinbib,floatfix]{revtex4-2}

\usepackage{graphicx}
\usepackage{bm}
\usepackage[dvipsnames]{xcolor}
\usepackage{tikz}
\usetikzlibrary{%
    decorations.pathreplacing,%
    decorations.pathmorphing,%
    arrows
}
\usetikzlibrary{positioning, calc,intersections,shapes,matrix,fit}

\usepackage{adjustbox}

\usepackage{amssymb}
\usepackage{enumerate}
\usepackage{comment}
\usepackage{listings}
\usepackage{mathrsfs}

\usepackage[colorlinks=true,linkcolor=MidnightBlue,citecolor=blue,filecolor=green,urlcolor=PineGreen]{hyperref}

\def\be{\begin{equation}}
\def\ee{\end{equation}}
\newcommand{\partl}[3]{ \frac{\partial^{#3}#1}{ \partial #2^{#3}} }	

\newcommand{\ket}[1]{\left\vert #1 \right\rangle}
\newcommand{\bra}[1]{\left\langle #1 \right\vert}

\newcommand{\tr}[1]{\mathrm{tr}\left( #1 \right)}

\definecolor{xkcd_purple}{RGB}{126, 30, 156}
\definecolor{xkcd_lilac}{RGB}{206, 162, 253}
\definecolor{xkcd_sky}{RGB}{117, 187, 253}
\definecolor{xkcd_cobalt}{RGB}{3, 10, 167}
\definecolor{xkcd_pine}{RGB}{10, 72, 30}
\definecolor{xkcd_teal}{RGB}{2, 147, 134}
\definecolor{xkcd_aquamarine}{RGB}{4, 216, 178}
\definecolor{xkcd_chartreuse}{RGB}{193, 248, 10}
\definecolor{xkcd_goldenrod}{RGB}{250, 194, 5}
\definecolor{xkcd_pumpkin}{RGB}{225, 119, 1}
\definecolor{xkcd_brightred}{RGB}{255, 0, 13}
\definecolor{xkcd_deeppink}{RGB}{203, 1, 98}
\definecolor{xkcd_wine}{RGB}{128, 1, 63}
\definecolor{xkcd_charcoal}{RGB}{52, 56, 55}
\definecolor{xkcd_greypurple}{RGB}{130, 109, 140}

\definecolor{codegreen}{rgb}{0,0.6,0}
\definecolor{codegray}{rgb}{0.5,0.5,0.5}
\definecolor{codepurple}{rgb}{0.58,0,0.82}
\definecolor{backcolour}{rgb}{0.95,0.95,0.95}
\definecolor{huntergreen}{rgb}{0.21, 0.37, 0.23}
\definecolor{lavenderpurple}{rgb}{0.59, 0.48, 0.71}
\definecolor{coquelicot}{rgb}{1.0, 0.22, 0.0}
\definecolor{crimsonglory}{rgb}{0.75, 0.0, 0.2}
\definecolor{deeppink}{rgb}{1.0, 0.08, 0.58}
\definecolor{electricviolet}{rgb}{0.56, 0.0, 1.0}
\definecolor{electricgreen}{rgb}{0.0, 1.0, 0.0}
\definecolor{mint}{rgb}{0.24, 0.71, 0.54}
\definecolor{dodgerblue}{rgb}{0.12, 0.56, 1.0}
\definecolor{lincolngreen}{rgb}{0.11, 0.35, 0.02}
\definecolor{persianblue}{rgb}{0.11, 0.22, 0.73}
\definecolor{pl}{rgb}{0.5, 0.0, 0.5}
\definecolor{amaranth}{rgb}{0.9, 0.17, 0.31}
\definecolor{candyapplered}{rgb}{1.0, 0.03, 0.0}
\definecolor{blueryb}{rgb}{0.01, 0.28, 1.0}

\begin{document}
\title{Pontryagin--Optimal Control of a non-Hermitian Qubit}

\author{Philippe Lewalle} \email{plewalle@berkeley.edu}
\affiliation{Department of Chemistry, University of California, Berkeley, CA 94720, USA}
\affiliation{Berkeley Center for Quantum Information and Computation, Berkeley, CA 94720, USA}
\author{K.~Birgitta Whaley} 
\affiliation{Department of Chemistry, University of California, Berkeley, CA 94720, USA}
\affiliation{Berkeley Center for Quantum Information and Computation, Berkeley, CA 94720, USA}

\date{\today}

\begin{abstract}
    Open--system quantum dynamics described by non-Hermitian effective Hamiltonians have become a subject of considerable interest. 
    Studies of non-Hermitian physics have revealed general principles, including relationships between the topology of the complex eigenvalue space and the breakdown of adiabatic control strategies.
    We study here the control of a single non-Hermitian qubit, similar to recently realized experimental systems in which the non-Hermiticity arises from an open spontaneous emission channel. 
    We review the topological features of the resulting non-Hermitian Hamiltonian and then present two distinct results. First, we illustrate how to realize any continuous and differentiable pure--state trajectory in the dynamics of a qubit that are conditioned on no emission. 
    This result implicitly provides a workaround for the breakdown of standard adiabatic following in such non-Hermitian systems. 
    Second, we use Pontryagin's maximum principle to derive optimal trajectories connecting boundary states on the Bloch sphere, using a cost function which balances the desired dynamics against the controller energy used to realize them. 
    We demonstrate that the latter approach can effectively find trajectories which maintain high state purity even in the case of inefficient detection. 
\end{abstract}

\maketitle


\section{Introduction} 
The dynamics of non-Hermitian (NH) systems have been an area of growing theoretical and experimental interest over the past two decades \cite{Bender_1998, Mostafazadeh_1, *Mostafazadeh_2, *Mostafazadeh_3, Heiss_2004, Heiss_2012, ElGanainy_2018, Ozdemir_2019, Miri_2019, Ashida_2020}.  
NH Hamiltonians arise naturally in the context of many open (non-conservative) systems. 
In contrast with Hermitian Hamiltonians, NH Hamiltonians may generically have complex eigenvalues and bi-orthogonal left and right eigenvectors. 
A number of unique effects can be understood in terms of the Riemann sheet topology of complex eigenvalues in parameter space; this topology is defined foremost by the presence of exceptional points (EPs), where there is a convergence of both
the complex eigenvalues and eigenvectors of the NH Hamiltonian \cite{Dembowski_2004, Zhong2018}. 
In particular, parameter loops that encircle EPs in NH systems may lead to dynamics exhibiting gain/loss effects that can break principles of adiabatic following \cite{Kvitsinsky_1991, Nenciu_1992, Berry_2011, Berry_2011_b, Leclerc_2013, Viennot_2014, Milburn2015} and manifest chiral behavior \cite{Doppler2016, Hassan2017, Zhong2018, Pick2019, Wang_2019, Zhong:19, Holler2020, Zhong_2021}.
There is now considerable literature about non-Hermitian systems in general, emphasizing both topological aspects of complex non-Hermitian spectra that may be understood statically \cite{Zhong2018, Holler2020}, and investigations of the corresponding dynamics \cite{Xu_2016, Lu_2018, WangClerck_2019, Wang_2021, Wang_2021_b}. 

One particular realization of NH physics arises in the conditional dynamics of open quantum systems \cite{Wu_2019, Naghiloo_2019_EP, Minganti_2019, Minganti_2020, Chen2021, Chen2022, abbasi2021topological}. 
Explorations of such systems have illustrated the applicability of many of the concepts from the classical context, and raised fundamentally new issues and potential applications. 
In this quantum context, two approaches to studying NH systems are apparent: topological properties of the NH Hamiltonian or Liouvillian \cite{Minganti_2019, Minganti_2020, Huber_2020} have been the subject of considerable interest, while the open and conditional quantum state dynamics underlying those NH properties may be directly understood via quantum trajectory theory \cite{BookCarmichael, BookBarchielli, BookWiseman, BookJacobs, Lewalle_2020Flor}. 
In the limit of slowly varying control parameters, the adiabatic principle suggests a clear connection between the system dynamics and NH topology. 
However, since the adiabatic principle is of limited applicability in NH systems~\cite{Nenciu_1992}, this slow limit does not offer a complete correspondence between the two, and connection of an NH system's dynamics to the underlying topology is not straightforward. 
For this fundamental reason and for practical considerations as well, the development of NH quantum control protocols that act effectively on shorter timescales is an important challenge for the field.
To this end, we note there have been recent proposals employing perturbative approaches \cite{Ribiero_2021}, NH shortcuts to adiabaticity \cite{Ibanez_2011, *Ibanez_2014, Torosov_2013,Chen_STA_2018}, as well as investigation of experimentally--realizable periodic controls \cite{abbasi2021topological}. 

\begin{figure}
    \centering
\includegraphics[width = \columnwidth]{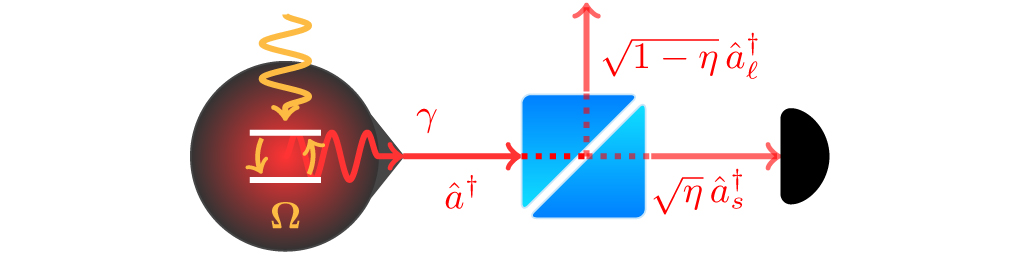}
    \caption{We sketch a possible realization of the scheme we consider. Unitary operations characterized by the Rabi rates $\boldsymbol{\Omega}$ are applied to a qubit. An effective non-Hermitian Hamiltonian arises due to the qubit's spontaneous emission. The dynamics conditioned on emission or non-emission of a photon are accessible if that emission is captured by a cavity or transmission line whose output is monitored by a photo--counter. Detection with efficiency $\eta$ can be modeled by imagining that a beamsplitter between the system and detector causes some photons to be lost. }
    \label{fig-device}
\end{figure}

We propose here an alternative approach, based on use of the Pontryagin maximum principle, which is ubiquitous both in the classical and quantum control theory literatures \cite{LiberzonBook, SchattlerBook, DAllessandroBook, Boscain2021}. 
We show that this can be usefully applied to the control of NH quantum systems, 
investigating in particular the behavior of Pontryagin control for a relatively simple example of a single qubit with an effective NH Hamiltonian arising from its spontaneous emission channel (see Fig.~\ref{fig-device}). 
This NH qubit system is similar to that studied in recent experiments \cite{Naghiloo_2019_EP, Chen2021, Chen2022, abbasi2021topological}, and as we show in this work, it proves to be a good testing ground for coherent control in the context of a NH quantum system with EPs. 

The rest of the paper is organized as follows.
We review the system dynamics, define the non-Hermitian Hamiltonian, and review its topological properties in Sec.~\ref{sec-NHdef}. 
In Sec.~\ref{sec-basic_control}, we take advantage of the relative simplicity of our chosen system, and demonstrate that it is straightforward to derive the controller paths needed to force the system along any continuous and differentiable pure state trajectory. 
In Sec.~\ref{sec-pontryagin} we formulate a cost function, and then apply Pontryagin's principle to derive the corresponding optimal trajectories the NH qubit can follow to connect boundary states over a desired evolution time. 
In particular, we illustrate that by using Rabi rotations as a control knob, arbitrary qubit state manipulations can be implemented under the NH dynamics, under both ideal (unit efficiency) detection of, or post--selection on, photon emission.
Furthermore, decoherence under partially--conditional dynamics (i.e.,~those arising from inefficient detection of emitted photons) can be substantially mitigated. 
In Sec.~\ref{sec-conclude} we provide a discussion of our results and potential future research based on the ideas we develop in this work.

\section{Non-Hermitian Jumpless Qubit Dynamics \label{sec-NHdef}}

{\color{black} We now consider the dynamics of a qubit that can spontaneously emit a photon into its environment with some probability, but \emph{does not actually emit}. 
These dynamics are accessible, e.g.,~if the qubit emits into a readout cavity and transmission line that route any photons to a photodetector.
The dynamics of interest are then those which occur between any detector clicks, i.e.~we focus on those dynamics which are inferred with knowledge that no photon was emitted through the open decay channel.
These dynamics are thus conditioned on the outcomes of the photon counting measurements.
This can also be realized in practice through post--selection using a third level \cite{Naghiloo_2019_EP}. 
The required continuous monitoring of the decay channel 
by photon counting \cite{Lewalle_2020Flor} is illustrated in Fig.~\ref{fig-device}, and described in greater detail in Appendix \ref{app-HeffDerivation}.
We include the measurement efficiency $\eta$, which allows us to consider imperfect photon counters, as well as to interpolate between the case where the detector is present and when it is absent.}

The conditional evolution of the qubit density matrix $\rho$ is given by
{\color{black} a modified Lindblad master equation}
\be \label{NH_Conditional}
\dot{\rho} = i\,\rho\,\hat{\mathcal{H}}^\dag - i \,\hat{\mathcal{H}}\,\rho+ (1-\eta)\hat{L}\,\rho\,\hat{L}^\dag + \eta\,\rho\,\tr{\hat{L}\,\rho\,\hat{L}^\dag},
\ee
where $\hat{L} = \sqrt{\gamma}\,\hat{\sigma}_-$ represents decay of the qubit into its environment at rate $\gamma$ {\color{black} (or, 
equivalently, with characteristic time $T_1 = \gamma^{-1}$)}, $\eta \in [0,1]$ is the photon detection efficiency, and 
\be \begin{split}
\hat{\mathcal{H}} &= \tfrac{1}{2}\left(\Omega_x\,\hat{\sigma}_x + \Omega_y\,\hat{\sigma}_y+\Omega_z\,\hat{\sigma}_z\right) - \tfrac{i}{2}\,\gamma\,\hat{\sigma}^+\hat{\sigma}_- \\
& = {\color{black} \frac{1}{2}\left( \begin{array}{cc}
\Omega_z - i\,\gamma & \Omega_x-i\,\Omega_y \\ 
\Omega_x+i\,\Omega_y & -\Omega_z
\end{array} \right)}
\end{split} \ee
is a non-Hermitian effective Hamiltonian that includes both unitary controls $\hat{H}  = \tfrac{1}{2}\,\boldsymbol{\Omega}\cdot\hat{\sigma}_\mathbf{q}$ and NH decay dynamics. 
{\color{black} Eq.~\eqref{NH_Conditional} gives the standard Lindblad master equation for $\eta = 0$ (no detection). 
We shall be most focused on the no--click decay dynamics accessible with ideal detection ($\eta = 1$), since the NH effective Hamiltonian $\hat{\mathcal{H}}$ is most closely associated with this ideal case.}
The dynamics \eqref{NH_Conditional} can equivalently be written as a dynamical system in the Bloch coordinates $\mathbf{q} = (x,y,z)^\top$ 
according to
$\dot{\mathbf{q}} = \tr{\hat{\sigma}_\mathbf{q}\,\dot{\rho}}$, where $\hat{\sigma}_\mathbf{q}$ are Pauli matrices. 
We will often refer to these dynamics using the shorthand notation $\dot{\mathbf{q}} = \boldsymbol{\mathcal{F}}(\mathbf{q},\boldsymbol{\Omega})$, 
noting that we may decompose the total dynamics into a sum of the jumpless NH dynamics ($\boldsymbol{\Omega} = 0$), plus the unitary part ($\boldsymbol{\Omega} \neq 0$), according to
\be \label{dotq_Fform}
\dot{\mathbf{q}} = \underbrace{\boldsymbol{\mathcal{F}}(\mathbf{q},\boldsymbol{\Omega} = 0)}_{\boldsymbol{\mathcal{F}}_0} + \boldsymbol{\Omega}\times\mathbf{q} = \boldsymbol{\mathcal{F}}(\mathbf{q},\boldsymbol{\Omega}). 
\ee

\begin{figure*}
    \centering
    \includegraphics[width = \textwidth]{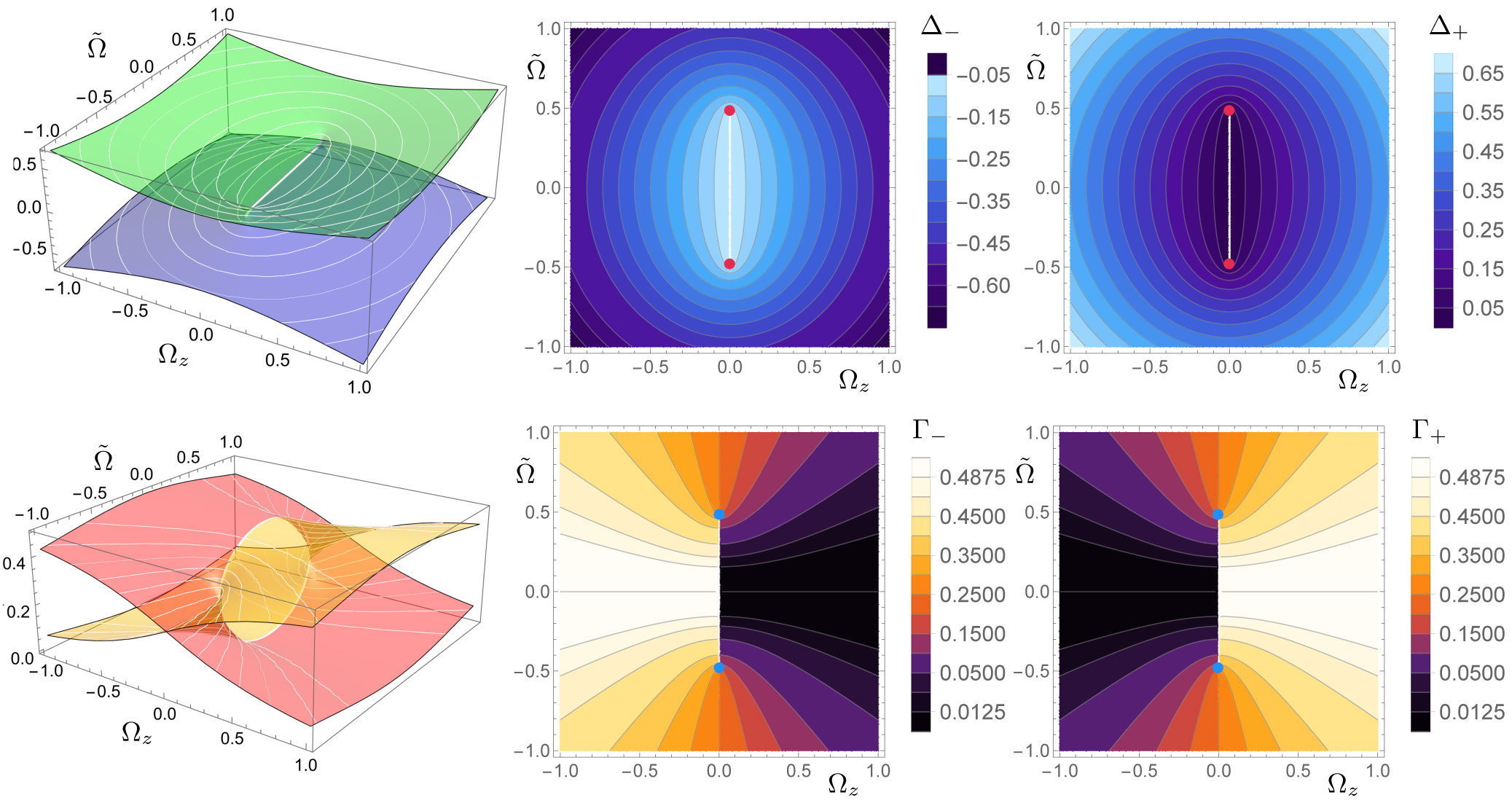}
    \caption{We plot the real parts of the right eigenvalues $\lambda_\pm$ of $\hat{\mathcal{H}}$ (i.e.~$\Delta_\pm$, top row), and the imaginary parts of $\lambda_\pm$ (i.e.~$\Gamma_\pm$, bottom row). 
    We fix $\gamma$, and all other rates ($\Omega$, $\Delta$, and $\Gamma$) are here expressed in units of $\gamma$. 
    The resulting pair of Riemann sheets appear together on the left, and then the real and imaginary parts of each individual sheet are shown on the right.
    We mark the pairs of EPs with markers $\bullet$, and choose to place our branch cut between the two EPs along the $\Omega_z = 0$ line. 
    Notice that a closed parameter loop enclosing one or both EPs ($n = 1,2$) will force $n$ exchange(s) between whether $\Gamma_+$ or $\Gamma_-$ is greater; this is a topological property of the complex spectrum, that is associated with the failure of adiabatic control protocols when encircling EPs. 
    This system differs from that of Ref.~\cite{Naghiloo_2019_EP} only in minor details. 
    }
    \label{fig_RSheets}
\end{figure*}

The right eigenvalues of $\hat{\mathcal{H}}$ read 
\be 
\lambda_\pm = -\tfrac{i}{4}\gamma \pm \tfrac{1}{4}\sqrt{4\boldsymbol{\Omega}^2-\gamma^2-4i\,\gamma\,\Omega_z} = -\tfrac{i}{4}\gamma \pm \tfrac{1}{4}\sqrt{\mathcal{J}}, 
\ee 
with corresponding right eigenvectors 
\be 
\ket{\lambda_\pm} = \mathcal{N}\left\lbrace \left(
2\Omega_z-i\gamma \pm\sqrt{\mathcal{J}}\right) \ket{e} + 2(\Omega_x+i\Omega_y) \ket{g} \right\rbrace
\ee 
(where $\mathcal{N}$ is a normalization factor). 
The left eigenvalues are the complex conjugates of the right eigenvalues. While $\ket{\lambda_+}$ and $\ket{\lambda_-}$ are no longer necessarily orthogonal to one another, the left and right eigenvectors do obey a bi-orthogonality relation \cite{Ashida_2020}. 
A pair of Exceptional Points (EPs) appear as the solutions with $\mathcal{J} = 0$, satisfying $\Omega_z = 0$ and $4(\Omega_x^2 + \Omega_y^2) = \gamma^2$. 
The EPs are located on the equator of the Bloch sphere, and manifest as fixed points there when the Rabi drive $\boldsymbol{\Omega}$ tries to excite the qubit at a rate that exactly balances the conditional decay dynamics $\gamma$ (e.g.~at $z = 0$ and $x = 1$, $\dot{z} = 0$ is achieved by $\Omega_y = -\gamma/2$, which lies on the ring of EPs defined by $\Omega_x^2+\Omega_y^2 = \gamma^2/4$). 
In the typical quantum--optical or circuit QED situations that are well--suited to realizing our physical situation, it is much more straightforward to modify $\boldsymbol{\Omega}$ precisely and quickly than to change $\gamma$.
We will consequently proceed by assuming $\gamma$ to be fixed, while treating $\boldsymbol{\Omega}$ as a dynamic control knob. 
Notice that the eigenvalues $\lambda_\pm$ have a rotational symmetry in $\Omega_x$ and $\Omega_y$; we may consequently define $\tilde{\Omega}$ to be a drive vector in the equatorial $xy$ Bloch--plane, and then understand the complex eigenspectrum by visualizing it as a function of $\tilde{\Omega}$ and $\Omega_z$. 
Fig.~\ref{fig_RSheets} shows the real and imaginary parts of $\lambda_\pm$ in this parameter space. 

Dynamics with a complex eigenspectrum differ qualitatively from those of Hermitian systems with real eigenvalues. 
Even a na\"{i}ve writing of the (un-normalized and non-trace--preserving) evolution 
$e^{-i \,\hat{\mathcal{H}}\,t} \sim \exp\left[ - i\,\lambda_\pm\,t \right] = \exp \left[-\Gamma_\pm\,t - i\,\Delta_\pm\,t \right]$ reveals that while the real parts of $\lambda_\pm$ lead to effective eigenenergies $\Delta \pm \equiv \Im[i\,\lambda_\pm]$, the imaginary parts $\Gamma_\pm \equiv \Re[i\,\lambda_\pm]$ behave as effective gain/loss parameters. 
Their significance becomes clear when considering the possibility of adiabatic control (e.g.,~by fixing $\gamma$ and then modifying the NH eigenstates by varying $\boldsymbol{\Omega}(t)$ slowly compared to $\gamma$). 
One finds that quasi-adiabatic following is possible, but only of the stable right eigenstate (where the more stable one at any given time is defined by having the smaller loss parameter) \cite{Nenciu_1992}.
In our current example we generically will have one right eigenstate closer to $\ket{e}$ and the other closer to $\ket{g}$ (this is true except at the EP); the right eigenstate closer to $\ket{g}$ will be the stable one, at the expense of the other.
Adiabatic following will be broken along any control trajectory that exchanges the stability of the two eigenstates.
Such a stability exchange is topologically guaranteed when performing closed--loop encircling of an EP \cite{Zhong2018}. 
As such, the difference between the relative losses $\Gamma_\pm$ along a parameter trajectory concisely explain many of the NH phenomena reported in the literature, including breaking adiabaticity \cite{Nenciu_1992}, and chiral state exchange \cite{Doppler2016, Hassan2017, Zhong2018, Pick2019, Wang_2019, Holler2020}. 
In our present example this means that adiabatic following works only in the lower half of the Bloch sphere (closer to $\ket{g}$), and will fail on a timescale $T_1$ if we try to switch hemispheres. 

The system \eqref{NH_Conditional} we have introduced is one of the simplest possible NH quantum systems one could devise. 
It is nevertheless surprisingly rich, containing spectral and dynamical features unique to NH systems, as well as being experimentally feasible \cite{Naghiloo_2019_EP}. 
This system is consequently an ideal sandbox for investigating NH physics and quantum control strategies \cite{Kvitsinsky_1991, Chen_STA_2018}. 

\section{Basic Control: Forcing the Qubit State along a Preset Trajectory \label{sec-basic_control}}

Above we describe two distinct pictures of the same system: 
On the one hand, we have a qubit with a complex eigenvalue spectrum containing exception points, and exhibiting a variety of behaviors that can be explained using the bi-orthogonal NH eigenstates. 
On the other hand, quantum trajectory theory tells us that the system dynamics may be straightforwardly expressed by the dynamical system \eqref{NH_Conditional} or \eqref{dotq_Fform} across timescales; this picture implicitly contains features of the NH topology, but can be applied without careful analysis of that complementary topological NH structure \cite{Lewalle_2020Flor}. 

\begin{figure}
    \centering
    \includegraphics[width = \columnwidth]{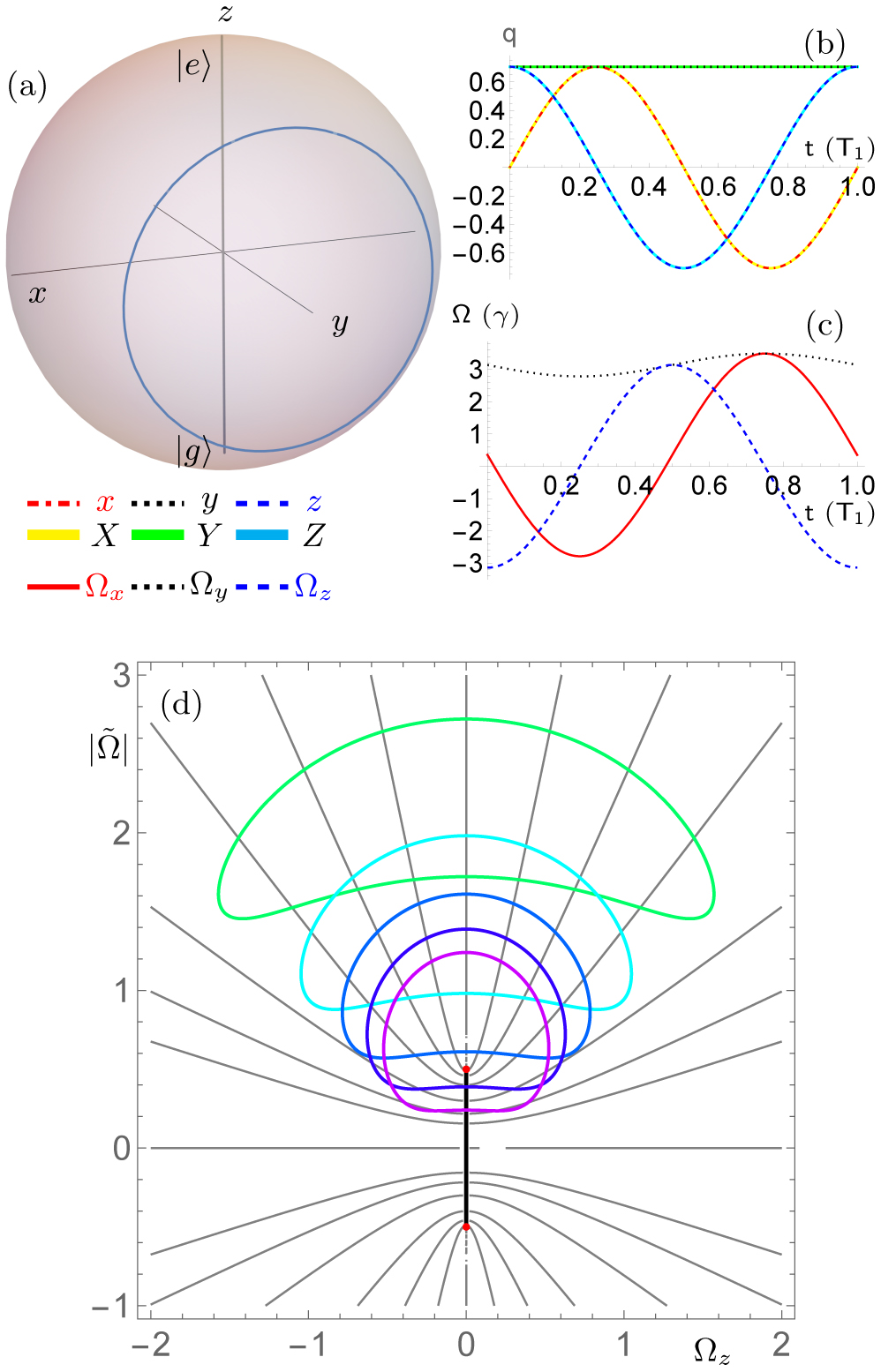}
    \caption{We illustrate a sample target trajectory $\mathbf{Q} = (X,Y,Z)^\top =  \tfrac{1}{\sqrt{2}}(\sin(2\pi\,t/T),1,\cos(2\pi\,t/T))^\top$ on the surface of the Bloch sphere in (a). We are able to exactly follow this contrived target trajectory (b) via a smooth controller trajectory (c), both of which are evaluated here for final time $T = 1\,[T_1]$. 
    In (d) we plot the contours of $\Gamma$ (gray), with the branch cut in black and EPs in red. 
    The $\boldsymbol{\Omega}(t)$ trajectories for $T = 2\,[T_1]$ (green), $T = 3\,[T_1]$ (cyan), $T = 4\,[T_1]$ (blue), $T = 5\,[T_1]$ (purple), and $T = 6\,[T_1]$ (magenta), are superposed over the contour plot. Note that to keep the figure two--dimensional, we summarize azimuthal unitary motion by the magnitude $|\tilde{\Omega}| = \sqrt{\Omega_x^2+\Omega_y^2}$. 
    We may understand the oscillation of $\Omega_y$ in (c) as correctly compensating for the changes in the gain loss ratio required to maintain the desired trajectory. 
    } \label{fig-RingY}
\end{figure}

To illustrate the power of the latter viewpoint, we show that we may straightforwardly force our qubit to follow any pure state trajectory we wish, assuming perfect conditional evolution {\color{black} (obtained with the measurement efficiency $\eta = 1$)}. 
In this case, our state $\mathbf{q}$ will be a unit vector pointing from the origin towards a point on the surface of the Bloch sphere, and $\boldsymbol{\Omega}$ will necessarily be an orthogonal vector generating rotations on the surface via the cross product in \eqref{dotq_Fform}. 
We have dynamics $\dot{\mathbf{q}} = \boldsymbol{\mathcal{F}}(\mathbf{q},\boldsymbol{\Omega} = 0) + \boldsymbol{\Omega}\times\mathbf{q}$; suppose we now solve for the drive $\boldsymbol{\Omega}$ that forces the qubit to follow a desired pure state target trajectory $\mathbf{Q}(t)$ on the surface. 
We assume that $\mathbf{Q}(t)$ is continuous and differentiable, such that $\dot{\mathbf{Q}}$ exists over the entire time domain of interest. 
For such a pure state trajectory, $\mathbf{q}$ will point to the surface of the Bloch sphere, while $\dot{\mathbf{Q}}$ will be a tangent vector along the surface of the sphere; it follows that $\boldsymbol{\Omega} = \Omega\,\bar{\boldsymbol{\Omega}}$ must be a mutually--orthogonal vector pointing along the direction
\be {\color{black}
\bar{\boldsymbol{\Omega}} = \frac{\mathbf{q}\times(\dot{\mathbf{Q}} - \boldsymbol{\mathcal{F}}_0)}{|\mathbf{q}\times(\dot{\mathbf{Q}} - \boldsymbol{\mathcal{F}}_0)|}\bigg|_{\mathbf{q}\,\rightarrow\,\mathbf{Q}},}
\ee
where we use the overbar to denote a unit vector, $\Omega = |\boldsymbol{\Omega}|$ is the magnitude, and $\boldsymbol{\mathcal{F}}_0$ is a shorthand for the conditional dynamics due to decay only. 
This expression suggests that we may implement a controlled unitary characterized by $\boldsymbol{\Omega}$ that ``makes up the difference'' between the conditional decay dynamics and {\color{black} the} target trajectory.
The magnitude can be obtained by rearranging $\dot{\mathbf{q}} = \boldsymbol{\mathcal{F}}_0 + \boldsymbol{\Omega}\times\mathbf{q} = \dot{\mathbf{Q}}$, and using the fact that $\mathbf{q}$, $\dot{\mathbf{Q}}-\boldsymbol{\mathcal{F}}_0$, and $\boldsymbol{\Omega}$ are all mutually orthogonal by construction. 
In summary then, we may drive the system along an arbitrary {\color{black} pure--state} trajectory $\mathbf{Q}(t)$ by applying the drive
\be \label{opt_omega_track}
{\color{black} \boldsymbol{\Omega} = \mathbf{Q}\times(\dot{\mathbf{Q}} - \boldsymbol{\mathcal{F}}_0(\mathbf{Q}))}
\ee
{\color{black} for the ideal $\eta = 1$ case.
An example of these dynamics is shown in Fig.~\ref{fig-RingY}.
Note that for the more general case of mixed states or $\eta < 1$, a generalization of Eq.~(7) as per $\boldsymbol{\Omega} = \mathbf{q} \times (\dot{\mathbf{Q}} - \boldsymbol{\mathcal{F}}_0^\perp)/|\mathbf{q}|^2$ may be used. Here $\boldsymbol{\mathcal{F}}_0^\perp$ is the component of $\boldsymbol{\mathcal{F}}_0 = \boldsymbol{\mathcal{F}}_0^\perp + \boldsymbol{\mathcal{F}}_0^{(R)}$ that is tangent to the sphere of radius $|\mathbf{q}|$; the remaining radial component $\boldsymbol{\mathcal{F}}_0^{(R)}$ cannot be directly cancelled by a unitary operation, with the result that $\mathbf{q}(t)$ is no longer constrained to follow the target trajectory $\mathbf{Q}(t)$ perfectly at all times. 
Note also that pinning the qubit to a desired pure state emerges naturally from this analysis as a simple sub-case of the dynamics under perfect measurement efficiency $\eta = 1$.}

Even though the example in Fig.~\ref{fig-RingY} includes a switch between hemispheres of the Bloch sphere, this trajectory does not necessarily imply a parameter loop that encloses an EP. 
{\color{black} We illustrate this in Fig.~\ref{fig-RingY}(d), with a set of trajectories corresponding to dynamics at differing speeds along the specific spatial trajectory shown in Fig.~\ref{fig-RingY}(a). 
It is evident that EP enclosure depends here on the speed at which we follow the path, with encircling occurring only for the slower loops.} {\color{black} Nevertheless, these different parameter trajectories all result in identical dynamics on the Bloch sphere over their respective time intervals (i.e., the path shown in panel (a) of Fig.~\ref{fig-RingY}).
While the EP and its attendant topology characterize the dynamics in the adiabatic limit (variations in $\boldsymbol{\Omega}$ much slower than $T_1$), other dynamical effects obscure those topological features when there is not an adiabatic--like separation of timescales. 
} 

Finally, we remark that the solutions \eqref{opt_omega_track} include the capability to realize {\color{black} something like a} ``shortcut to adiabaticity'', even in the case where the actual adiabatic following fails, in {\color{black} this sense:} 
First, {\color{black} we may define} any parameter trajectory $\boldsymbol{\Omega}_0$ and then find the Bloch trajectory $\mathbf{Q}_0(t)$ which corresponds to the motion of a NH eigenstate under $\boldsymbol{\Omega}_0$. We can then use \eqref{opt_omega_track} to compute a new parameter trajectory $\boldsymbol{\Omega}_1$ which drives the qubit along $\mathbf{Q}_0(t)$ over any timescale.\footnote{Since the NH eigenstate depends on $\mathbf{\Omega}$, we will not necessarily be following the real--time NH eigenstate of $\boldsymbol{\Omega}_1$; we will just be following the NH eigenstate path that was defined by the other parameter path $\boldsymbol{\Omega}_0$. This will be the case for any control additions which modify the eigenstates of the Hamiltonian.} 
Similarly, one could also use \eqref{opt_omega_track} to accelerate the NH dynamics that the system exhibits in response to a slowly--varying $\boldsymbol{\Omega}$ (slow enough to have adiabatic motion plus modifications primarily due to the NH topology described in the previous section).


\section{Pontryagin Optimal Control: Cost Functions and Extremization \label{sec-pontryagin}}

We have demonstrated above that we can easily force our system to follow particular dynamics.
We now change emphasis, and consider controls that follow optimal dynamics between arbitrary boundary states with respect to some cost function, which we derive via the Pontryagin action extremization method. 
In other words, instead of specifying a target trajectory $\mathbf{Q}(t)$, we will now only specify an initial $\mathbf{q}_i$ and a final target state $\mathbf{Q}_f$, and derive optimal dynamics which connect them.  

Pontryagin optimal control \cite{DAllessandroBook, Boscain2021} assumes that one has 1) some dynamical equations $\dot{\mathbf{q}} = \boldsymbol{\mathcal{F}}(\mathbf{q},\boldsymbol{\Omega})$ encoding the system dynamics and their relationship to any available control knobs (here the Rabi drive $\boldsymbol{\Omega}$ takes on the latter role), and 2) some cost function $\mathscr{S}$ resembling an action functional. 
The aim is then to derive controller trajectories that extremize the cost $\mathscr{S}$ while also satisfying the dynamical constraint $\boldsymbol{\mathcal{F}}$. 
As we have a dynamical equation \eqref{dotq_Fform}, we need only specify the cost against which we want to optimize. 
Generically, we wish to find controller trajectories $\boldsymbol{\Omega}(t)$ that take us rapidly from some initial state $\mathbf{q}_i$ to a chosen final state $\mathbf{Q}_f$, while using our drive power $\boldsymbol{\Omega}^2$ efficiently. 
We may consequently write a cost function of the form
\begin{subequations} \be 
\mathscr{S} = 
\int_0^T dt\,\left(\boldsymbol{\Lambda}\cdot\dot{\mathbf{q}} - \mathscr{H} \right),
\ee
where the Hamiltonian 
\be \label{H-opt}
\mathscr{H} = \boldsymbol{\Lambda}\cdot\boldsymbol{\mathcal{F}}(\mathbf{q},\boldsymbol{\Omega}) - \tfrac{\alpha}{2}(\mathbf{q} - \mathbf{Q}_f)^2 - \tfrac{\beta}{2}\boldsymbol{\Omega}^2
\ee\label{eq:costs}\end{subequations} 
includes the Lagrange multipliers (co-states) $\boldsymbol{\Lambda}$ that constrain solutions to the dynamics of interest, as well as penalties for being far away from the target final state and for using high drive power. 
The co-states are not physical; they enter only as a computational tool to derive the Pontryagin--optimized dynamics. 
{\color{black} We remark} that under no circumstances should the object $\mathscr{H}$ used for optimization be confused with the quantum operators $\hat{H} = \hat{H}^\dag$ or $\hat{\mathcal{H}} \neq \hat{\mathcal{H}}^\dag$. 
The coefficients $\alpha$ and $\beta$ are introduced in \eqref{H-opt} so that we may tune the relative weights of the different cost terms that we choose to include.  

Pontryagin's principle then manifests as action extremization, i.e.,~taking $\delta \mathscr{S} = 0$ leads to 
\begin{subequations} \label{optimal-eqs} \be
\partial_{\boldsymbol{\Lambda}} \mathscr{H} = \dot{\mathbf{q}} = \boldsymbol{\mathcal{F}}, \ee
\be \partial_{\mathbf{q}} \mathscr{H} = -\dot{\boldsymbol{\Lambda}}, \ee \be \partial_{\boldsymbol{\Omega}}\mathscr{H}|_{\boldsymbol{\Omega} = \boldsymbol{\Omega}^\star} = 0,
\ee \end{subequations}
which are necessary conditions for optimality. 
We have Hamilton's equations of motion, plus a condition for deriving optimal controller trajectories $\boldsymbol{\Omega}^\star(t)$. 
{\color{black} Details of the optimal dynamics appear in Appendix \ref{app-pontryagin}.}
It is straightforward to solve this last equation and find that optimal controller trajectories obey
$\boldsymbol{\Omega}^\star = \tfrac{1}{\beta} \,\mathbf{q} \times \boldsymbol{\Lambda}$. 
Plugging in this optimal Rabi drive $\boldsymbol{\Omega}$ then yields the Hamiltonian 
\be 
\mathscr{H}^\star =  \tfrac{1}{2\beta}\left\lbrace \mathbf{q}^2\,\boldsymbol{\Lambda}^2 - (\mathbf{q}\cdot\mathbf{\Lambda})^2\right\rbrace + \boldsymbol{\Lambda}\cdot\boldsymbol{\mathcal{F}}_0- \tfrac{\alpha}{2}(\mathbf{q}-\mathbf{Q}_f)^2. 
\ee

The solutions of Hamilton's dynamical equations generated by $\mathscr{H}^\star$, initialized for a particular initial $\mathbf{q}_i$ and all possible $\boldsymbol{\Lambda}_i$, form a Lagrangian Manifold (LM) of candidate solutions \cite{BookArnoldClassical}; if a globally optimal solution for a given final boundary conditions exists, it can be found in this solution manifold. 
Note that for constant $\alpha$, $\beta$, and $\gamma$, an optimal solution necessarily conserves its Pontryagin energy $\mathscr{E} = \mathscr{H}$.
{\color{black} The main challenge in solving the optimal dynamics eq.~\eqref{optimal-eqs} is in finding solutions that match the particular boundary conditions $\mathbf{q}_i$ and $\mathbf{q}(T)$. 
That boundary value problem in the coordinate space can equivalently be formulated as an initial value problem in terms of the initial values of the states $\mathbf{q}_i$ and the co-states $\boldsymbol{\Lambda}_i$. 
We shall describe solutions to \eqref{optimal-eqs} on a Lagrangian Manifold (LM) that is defined from a subset of initial co-states $\boldsymbol{\Lambda}_i$ for given initial $\mathbf{q}_i$. 
Specifically, the LM containing all $\boldsymbol{\Lambda}_i$ contains all possible optimal solutions originating at the given $\mathbf{q}_i$. 
This LM can be understood as a tool to translate between the initial value formulation of the control problem and the boundary value formulation, because there will always be at least one initial co-state $\boldsymbol{\Lambda}_i$ that generates a path on the LM leading to any attainable $\mathbf{q}(T)$. 
The LM thus provides a way of understanding the mapping between the initial co-states and attainable final states at later times. 
This tool has previously been used in a similar way to describe quantum trajectories following an optimal readout \cite{Lewalle_Multipath, Naghiloo_Caustic, Lewalle_Chaos}, and is discussed further in Appendix \ref{app-pontryagin}.
}

The {\color{black} values of the }coefficients $\alpha$ and $\beta$ do impact the types of solutions {\color{black} that} we can get. 
In Appendix \ref{app-pontryagin} we show that $\alpha$ effectively determines how fast the co-states $\boldsymbol{\Lambda}$ change as a function of the distance from the target state. 
The form of the optimal Rabi drive clearly indicates that $\beta$ effectively scales how the co-states $\boldsymbol{\Lambda}$ translate to a controller trajectory, i.e.,~$\beta$ just re-scales the strength of the controlled unitary dynamics relative to the decay dynamics of $\boldsymbol{\mathcal{F}}_0$. 
In less precise but more suggestive terms, a larger value of $\alpha$ allows the controlled state trajectory to accelerate faster, while a large $\beta$ tempers the impact of erratic co-state dynamics on the actual controller dynamics. 
We shall consequently occasionally refer to $\alpha$ and $\beta$ as the controller's ``agility'' and ``temperance'', respectively, with an understanding that these parameters can partially counterbalance each other in practice.
For simplicity we will typically set the temperance $\beta = 1$ in the examples below. This is because a constraint on the maximum available $\mathscr{E}$ plays a similar role and we necessarily already have to limit that energy in numerical simulations by choosing a finite volume of initial co-states $\boldsymbol{\Lambda}$. 

\subsection{Pure State Control} 
\begin{figure}
\centering
\includegraphics[width = \columnwidth]{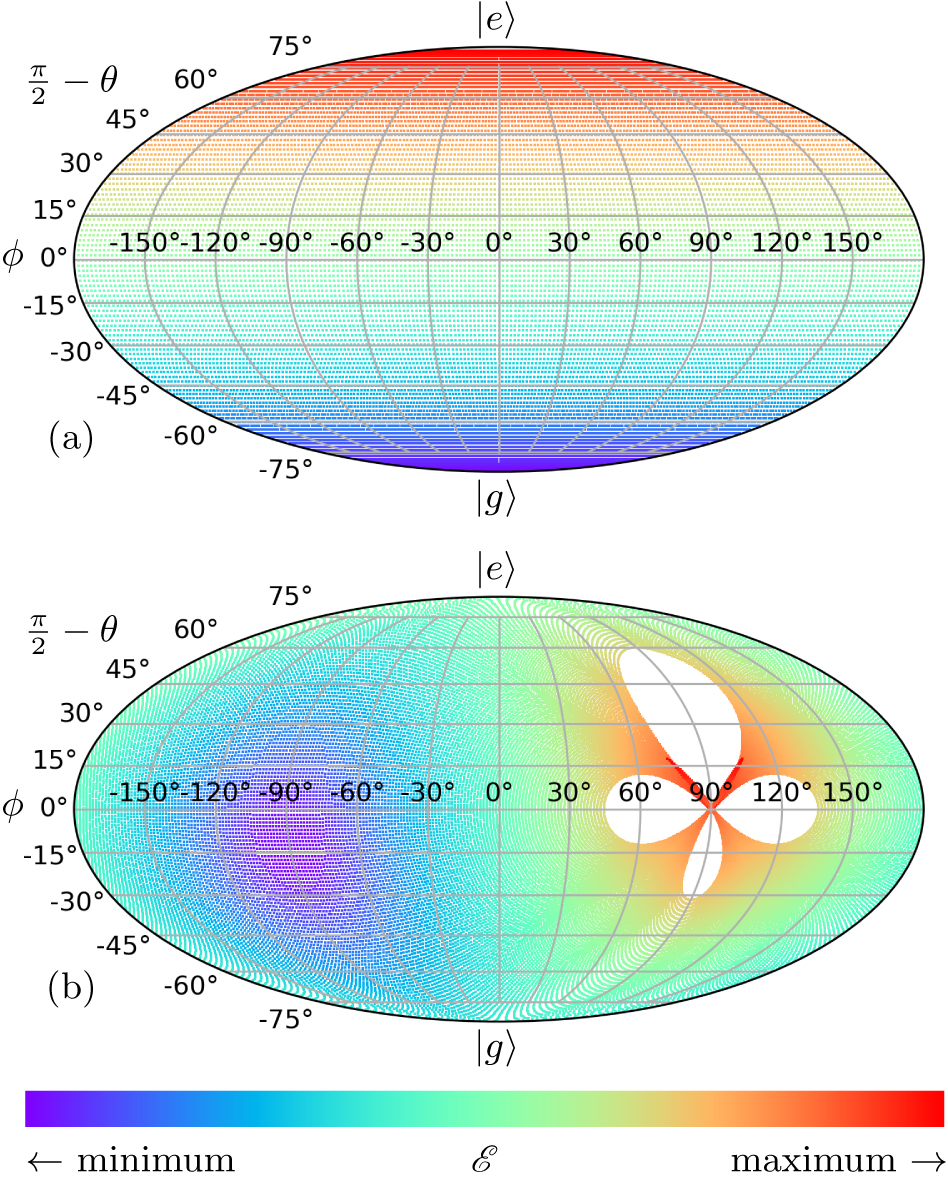}
    \caption{{\color{black} We plot samplings of the Lagrangian Manifold (LM) in two different situations. Each point in a sampled LM represents the state at time $T$ that is generated by a different initial co-state $\boldsymbol{\Lambda}_i$, such that the resulting plot of the sampled LM shows the extent of the possible optimal solutions on their way to the target state $\mathbf{Q}_f$.} 
The surface of the Bloch sphere is represented via Mollweide projection in both plots.
    {\color{black} (a) Plot of the LM after time $T = 2/3\,[\gamma]$, initialized at $\ket{g}$}, with $\eta = 1$, $\alpha = 2$, $\beta = 1$, and target state $\ket{e}$. 
    Panel (a) implies that if an unwanted jump to $\ket{g}$ were detected, there exist optimal solutions to reset the state in the next interval of no-jump evolution over a relatively short time and using modest control drive parameters. (b) Plot of the LM after $T = 1/2\,[\gamma]$, from a state initialized at $y = -1$, with dynamics targeting the orthogonal state $y = +1$.
    In both cases we are able to drive the transition of interest in reasonable times and find complete coverage of the Bloch sphere soon after the time shown, 
    {\color{black} indicating that there is a controller path reaching every pure final state under the conditions shown (even on the way to a different target state). Smaller values of $\mathscr{E}$ (bluer regions) generically correspond to control trajectories with smaller $|\boldsymbol{\Lambda}|$, i.e.,~trajectories with low $\mathscr{E}$ tend to be the slower ones that will reach a target state over longer time intervals. 
    Conversely, larger $\mathscr{E}$ values (redder regions) correspond to the fastest control paths that we allow, sitting on the leading edge of the manifold.
    The LM is bounded by a range of initial conjugate momenta (co-states) expressed in spherical coordinates, specifically $\Lambda_\theta(0) \in [-5,5]$ and $\Lambda_\phi(0) \in [-5,5]$ for both panels above. 
    Expansion of this initial $\boldsymbol{\Lambda}$--volume would reveal faster coverage of the Bloch sphere and solutions over shorter time intervals (at the expense of higher Rabi power).}
    Animations of the evolving LMs appear in the supplemental materials, {\color{black} and are described in Appendix \ref{app-SM-list}. The context of the snapshots above is best understood via such animations.} 
    }
    \label{fig_pure_control}
\end{figure}

We demonstrate {\color{black} here} that it is possible to solve the control problem framed above numerically in the case $\eta = 1$, and thereby obtain controller and dynamical solutions which map arbitrary initial qubit states $\mathbf{q}_i$ to arbitrary final qubit states $\mathbf{Q}_f$ on a desired timeframe, {\color{black} (i.e., in a time $T$)} within the context of the NH dynamics \eqref{NH_Conditional}. 
Fig.~\ref{fig_pure_control} illustrates two pertinent examples. 
Panel (a) shows that we can drive a transition from $\ket{g}$ to $\ket{e}$ against decay, and panel (b) shows that we can drive a transition between two orthogonal states on the equator of the Bloch sphere. Both transitions can be driven on timescales faster than the average spontaneous emission time $\gamma^{-1}$ with modest drive power.
We furthermore find that for both initial conditions the associated LM covers the entire Bloch sphere {\color{black} ``on its way'' to the target state}, indicating that for sufficiently large $\boldsymbol{\Omega}^2$ or $|\boldsymbol{\Omega}|$, i.e., for sufficiently fast Rabi driving, 
there are no forbidden pure states {\color{black} that} our control scheme cannot access. 

Assessing the examples of Figs.~\ref{fig-RingY} and \ref{fig_pure_control}, and the underlying methods, we arrive at the following conclusions. 
Despite the apparent complexity of the present system when viewed through the lens of a non-Hermitian Hamiltonian, there are actually no substantial barriers impeding pure--state control of the ideal ($\eta = 1$) system state if we approach it instead as a dynamical system. 
Particularly, we have in this case been able to solve for the drive that forces the system along an exact desired pure state trajectory, and further shown that we may formulate reasonable cost functions against which to derive optimized state dynamics and the controls that generate them. 
We have done this with a relatively simple cost function, but nothing prevents us from investigating other cost functions that are motivated by more specific tasks. 

\begin{figure}
\centering
\includegraphics[width = \columnwidth]{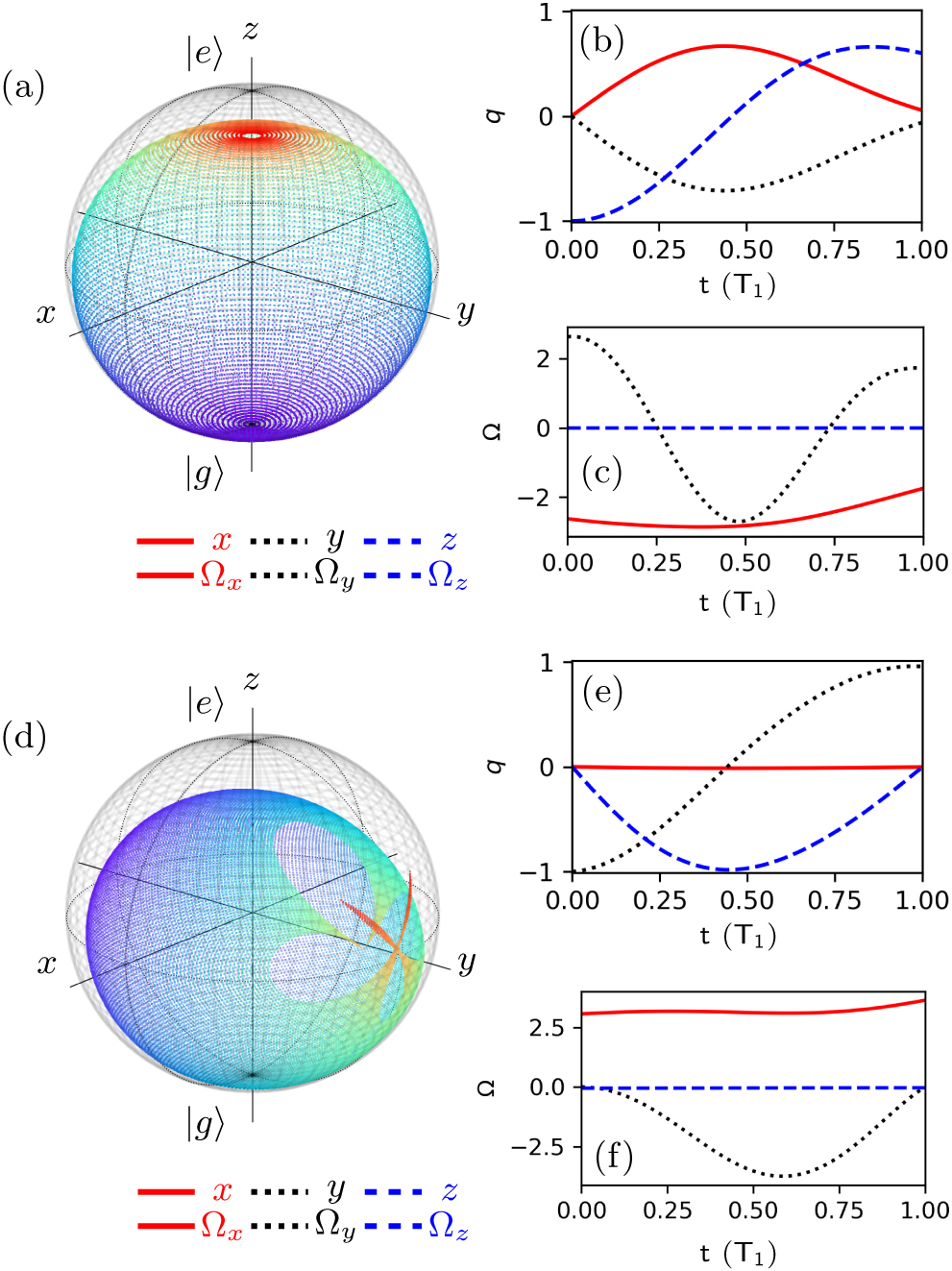}
\caption{
Panels (a) and (d) repeat the manifold integrations of Fig.~\ref{fig_pure_control}(a,b), respectively, using all the same parameters \emph{except} that we now have inefficient conditioning, $\eta = 0.5$. 
Mixing due to the now partially--monitored decay  is evident primarily close to $\ket{e}$, making pure states in the vicinity of $\ket{e}$ the most difficult to reach with modest values of $|\boldsymbol{\Omega}|$. 
Panels (b) and (c) show the optimally--controlled trajectory which comes closest to the target state $\ket{e}$ at $T = 1\,[T_1]$, and the controller trajectory $\boldsymbol{\Omega}(t)$ generating it. 
It is evident that this trajectory struggles to terminate at $z \gtrsim 0.5$, highlighting the difficulty in reaching the excited state when inefficient photodetection is used. 
We {\color{black} will} show in Fig.~\ref{fig-pontryagin-accel} {\color{black} below} that this effect can be mitigated by allowing for higher Pontryagin energy and greater controller agility $\alpha$. 
Panels (e) and (f) show the state trajectory and controller trajectory moving from $\ket{-y}$ to $\ket{y}$ over a duration $T = T_1$. 
Panel (e) shows a relatively high--fidelity operation, in stark contrast with panel (b). This is because we select here boundary conditions which can be reached by a trajectory which spends much of its time near $\ket{g}$, where the purity loss due to decay is relatively small. Further details appear in Appendix \ref{app-pontryagin}.
Animations of the evolving LMs appear in the supplemental materials, {\color{black} and are described in Appendix \ref{app-SM-list}.}
}
\label{fig-Pontryagin-mixed}
\end{figure} 

\begin{figure}
    \centering
    \includegraphics[width = \columnwidth]{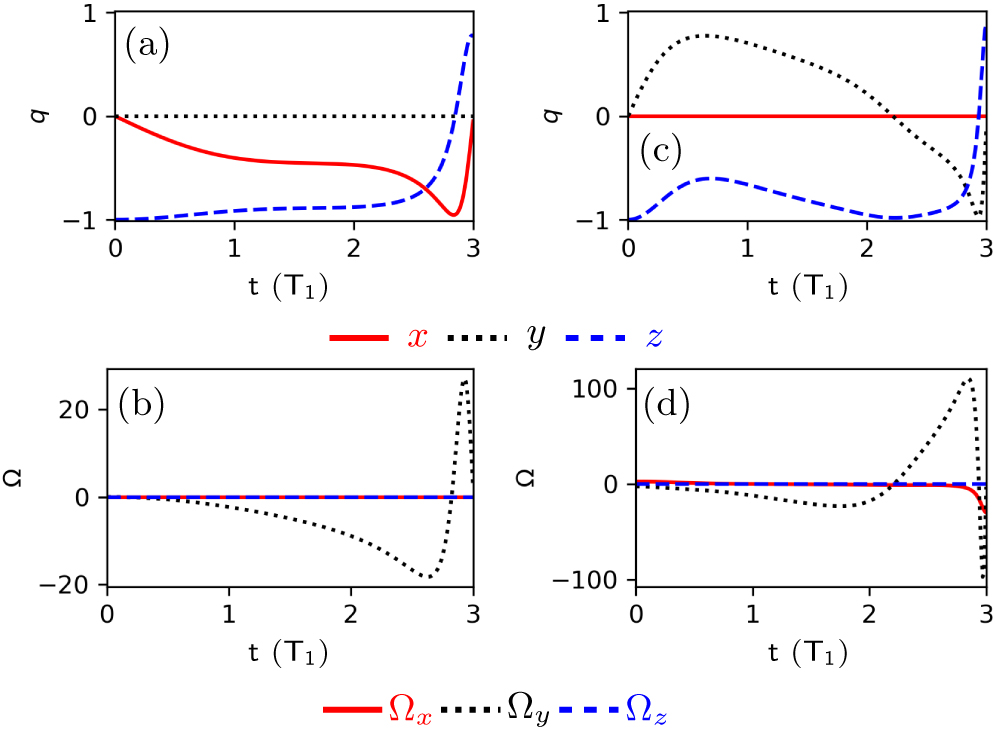}
    \caption{We repeat Fig.~\ref{fig-Pontryagin-mixed}(b,c), this time with an expanded LM $\Lambda_\theta \in [-10,10]$ and $\Lambda_\phi \in [-10,10]$, completely unconditional dynamics $\eta = 0$, and a longer duration $t = 3\,T_1$. 
    Panels (a,b) use $\alpha = 2$ in (a,b), and panels (c,d) increase the agility to $\alpha = 10$. (The temperance is $\beta = 1$ in all cases).
   It is evident that despite the significantly worsened (now non-existent) detection, and longer evolution time compared with Fig.~\ref{fig-Pontryagin-mixed}, we can actually get closer to $\ket{e}$. By allowing the controller use of greater power and the latitude to make more rapid adjustments, the optimization scheme is then able to keep a pure state near $\ket{g}$ until accelerating rapidly towards $\ket{e}$ at the end. Thus, the larger initial manifold and $\alpha$ values allow our controller to find trajectories in which decoherence is meaningfully mitigated. Further details appear in Appendix \ref{app-pontryagin}, and in Fig.~\ref{fig-PurDot}. }
    \label{fig-pontryagin-accel}
\end{figure}

\subsection{Mixed State Control and Inefficient Conditioning}

We continue by considering the less straightforward case with inefficient photon detection, $\eta < 1$. 
Inefficient detection or post--selection implies a loss of information that dissipates into environmental degrees of freedom without being detected and generically leads to mixed quantum states \cite{Lewalle_2020Flor}. 
This presents a clear challenge for control \cite{Koch_2016, Patti_2017}. 
In particular, many high--purity qubit states may not be accessible to moderate--time control trajectories, because unitary controls cannot directly ``undo'' decoherence caused by inefficient detection (i.e.,~an observer's control abilities are limited if they possess only incomplete knowledge about the system of interest). 
This is apparent in Fig.~\ref{fig-Pontryagin-mixed}, where we repeat the examples of Fig.~\ref{fig_pure_control} but with an imperfect detector efficiency $\eta = 0.5$.
A decay channel in particular will tend to mix the qubit state over short times if we are near $\ket{e}$, but will then re-purify the qubit over timescales much longer than $T_1$ without a drive (i.e.,~$\ket{g}$ is a pure state, and the system will either jump there or asymptotically approach it if left alone). 
We can consequently begin to qualitatively understand the route towards reaching high--purity target states with poor detection efficiency. 
Here, in addition to elements of the ideal analysis above, there ought to be an implicit benefit to control trajectories which go through the ground state (or near it), so as to minimize decoherence over the course of the evolution. 

To that end, an ideal trajectory to reach $\ket{e}$, which is the most difficult state to reach, is one that stays near $\ket{g}$ to retain purity for a relatively long time during its evolution, and then rapidly accelerates towards $\ket{e}$ at the last possible moment. 
We elaborate on this point in Appendix \ref{app-pontryagin}.
It is natural to ask whether our optimization scheme can actually come up with a solution for the trajectory just described. The answer is that it can, and such a solution is explicitly shown in Fig.~\ref{fig-pontryagin-accel}. 
The coefficient $\alpha$ turns out to play a meaningful role at this point: Increasing the ``agility'' allows the controller to change direction faster (see Appendix \ref{app-pontryagin}), which is helpful in minimizing decoherence. 
{\color{black} In particular, Fig.~\ref{fig-pontryagin-accel} shows that the difficulties in performing the operation $\ket{g}\rightarrow\ket{e}$ illustrated in Fig.~\ref{fig-Pontryagin-mixed}(a,b,c) can largely be overcome by increasing $\alpha$.}

\section{Discussion and Outlook \label{sec-conclude}} 

We have demonstrated here an instance where optimal control theory may be successfully applied to a non-Hermitian quantum system. 
We have focused on a relatively simple system (a decaying qubit) whose dynamics can be completely understood via quantum trajectory theory (see \cite{Lewalle_2020Flor} and references therein), and whose pure--state dynamics are straightforwardly controllable. 
This has allowed us to analyze the conceptual aspects of non-Hermitian physics relative to our control schemes from a strong foundation. 
We have used this system to review a number of phenomena that are generic to non-Hermitian Hamiltonians in quantum mechanics (exceptional points, bi-orthogonal eigenstates, and so on), and then demonstrated that complementary and powerful control methods arise straightforwardly when the same system is approached from a quantum trajectory perspective instead.  

Our results here constitute a particular type of feedback control \cite{Zhang_feedback}. 
The construction of a non-Hermitian Hamiltonian is conditioned on the readout (i.e.,~on the absence of photon emission events), but retains many of the desirable features of open--loop control between emission events.
In particular, controls for extended times between jumps may be pre-computed. 
We might consequently characterize our work above as an exploration of ``open--loop conditional quantum control''. Despite the importance of the measurement record, we do not require feedback to be computed instantaneously, as occurs in many feedback problems requiring non-differentiable controller trajectories that respond to diffusive conditional dynamics at each time interval \cite{BookWiseman, BookJacobs, Patti_2017, Zhang_feedback, Song_PAQS}. 

We further note that conditional evolution of an open quantum system often admits a description in terms of a NH Hamiltonian, such that general features of our approach may apply to a wide class of related problems. 
In particular we have developed a pathway towards mitigating the impacts of inefficient detection in this context, with demonstrated efficacy in a simple example. 
Our present work further suggests novel research avenues to consider for control of more complex non-Hermitian quantum systems, by probing the intersection between dynamical and topological pictures of the system. 

Much of the present interest in NH physics revolves around topological aspects of the spectrum \cite{Xu_2016, Holler2020, Patil_2021}. 
Such properties may be evaluated with respect to NH eigenvalue Riemann sheets as parameters are varied (as in, for example,~Fig.~\ref{fig_RSheets}), but connecting these ideas to the actual dynamical response of the system across arbitrary timescales and in general remains a challenge for the field. 
We view our application of Pontryagin's principle to a NH quantum system as one of many emerging strategies for the control of NH physics, which seek to avoid the limitations of adiabatic following in this context. 
The approach presented here is complementary to other efforts addressing non-Hermitian quantum control, including experimental work \cite{abbasi2021topological}, and theoretical studies based on perturbation theories \cite{Ribiero_2021} or shortcuts to adiabaticity \cite{Ibanez_2011, *Ibanez_2014, Torosov_2013, Li_STA_2017, Chen_STA_2018, Guery_RMP_2019}. 
Further development of a Pontryagin control approach for NH systems, and unification of topological NH concepts with a dynamical picture due to quantum trajectory theory, are promising avenues for continued research. 
Ultimately, we see this work as one element of a toolbox which may, together with other methods, lead to comprehensive control of non-Hermitian quantum systems that takes advantage of their unique physics for tasks of interest to quantum information science. 

\par \emph{Acknowledgements --- } This work has been supported by AFOSR MURI Grant No.~FA9550-21-1-0202. 
We are grateful to Rob Cook, Zengzhao Li, and the groups of Kater Murch and Jack Harris for insightful discussions about non-Hermitian physics related to the present work.   


\begin{center}
\end{center}
\appendix

\section{No--Jump Qubit Dynamics with a Decay Channel \label{app-HeffDerivation}}


We derive here our non-Hermitian Hamiltonian and the associated conditional dynamics, using a Bayesian description of continuously--monitored sponataneous emission (see Refs.~\cite{Lewalle_2020Flor, Jordan_2015Flor} and references therein). 
Physically, we assume that a qubit emits into a readout cavity (i.e.,~a fast decaying cavity) and/or transmission line, where the values of the qubit transition frequency and the density of environmental states result in emission at a rate $\gamma$ (or timescale $T_1 = 1/\gamma$), with the characteristic exponential decay of the excited state population (i.e.,~we assume that the types of approximations first made by Weisskopf and Wigner \cite{Weisskopf1930} apply). 
In this setting, emitted photons can be captured with high efficiency and then be routed to a detection device, such that the conditional evolution is accessible. See Fig.~\ref{fig-device}.  

It is then phenomenologically appropriate to consider a separable initial state of a qubit and output line leading to a detector, namely $(\zeta\ket{e}+\upsilon\ket{g})\otimes\ket{0}$, evolving to 
\begin{subequations}\be 
\zeta\left(\sqrt{e^{-\gamma\,\Delta t}}\ket{e,0}+\sqrt{1-e^{-\gamma\,\Delta t}}\ket{g,1}\right) + \upsilon \ket{g,0}
\ee
after a time $\Delta t$. This can equivalently be written as
\be \label{kraus-ideal}
\left( \begin{array}{cc}
\sqrt{e^{-\gamma\,\Delta t}} & 0 \\ \sqrt{1-e^{-\gamma\,\Delta t}}\,\hat{a}^\dag & 1
\end{array} \right) \left(\begin{array}{c} \zeta \\ \upsilon \end{array}\right) \otimes \ket{0},
\ee\end{subequations}
where $\hat{a}^\dag\ket{0} = \ket{1}$ represents the emission of a photon, and the matrix form highlights the action on the qubit state. 
Kraus operators can be obtained by selecting a final state of the line, which we will associate with a measurement outcome, leaving behind an operation on the qubit state only. 
We introduce a notion of loss between qubit and detector via insertion of a beamsplitter relation $\hat{a}^\dag \rightarrow \sqrt{\eta}\,\hat{a}^\dag_s + \sqrt{1-\eta}\,\hat{a}^\dag_\ell$, which states that an emitted photon may go to a monitored signal port $\hat{a}^\dag_s$ with probability $\eta$, or may be lost in transit with probability $1-\eta$. We will henceforth refer to $\eta$ as the measurement efficiency. 
This process generalizes \eqref{kraus-ideal} to
\be 
\underbrace{\left( \begin{array}{cc}
\sqrt{e^{-\gamma\,\Delta t}} & 0 \\ \sqrt{1-e^{-\gamma\,\Delta t}}\,(\sqrt{\eta}\,\hat{a}_s^\dag + \sqrt{1-\eta}\,\hat{a}_\ell^\dag ) & 1
\end{array} \right)}_{\mathcal{M}} \left(\begin{array}{c} \zeta \\ \upsilon \end{array}\right) \otimes \ket{00}. 
\ee
Note that the existing similar experiments emphasizing NH dynamics use a third qudit level and post--selection \cite{Naghiloo_2019_EP, Chen2021, Chen2022, abbasi2021topological} instead of continuous photo--counting, for practical reasons. 
This is however conceptually the same as the picture laid out here, since post--selection on no jumps having occurred to a third level at some time is equivalent to conditioning on an emission event not having occurred in any of the timesteps leading up to that moment. 

We now adapt these expressions to describe the conditional evolution of the qubit. 
Three distinct outcomes are possible for a measurement over a finite timestep $\Delta t$, each represented by a Kraus operator: The detector may click due to photon emission, as represented by the operator
\begin{subequations} \be \begin{split}
\hat{\mathcal{M}}_{10} &= \bra{1_s\,0_\ell} \mathcal{M} \ket{0\,0} = \left( \begin{array}{cc} 0 & 0 \\ \sqrt{\eta(1-e^{-\gamma\,\Delta t})} & {\color{black} 0} \end{array} \right) 
\\ &\approx \sqrt{\Delta t}\,\sqrt{\eta\,\gamma}\,\hat{\sigma}_- + \mathcal{O}\left( \Delta t^\frac{3}{2} \right),
\end{split} \ee 
or the detector may not click, which is due to either no photon being emitted 
\be \begin{split}
\hat{\mathcal{M}}_{00} &= \bra{0_s\,0_\ell}\mathcal{M}\ket{0\,0} = \left( \begin{array}{cc} \sqrt{e^{-\gamma\,\Delta t}} & 0 \\ 0 & 1 \end{array} \right) 
\\ &\approx \hat{\mathbb{I}} + \Delta t \underbrace{\left( \begin{array}{cc} -\gamma/2 & 0 \\ 0 & 0 \end{array} \right)}_{\hat{\mathfrak{Z}}} + \mathcal{O}\left(\Delta t^2\right), \\ &\quad \\ & \quad
\end{split} \ee
or to a photon being emitted but going unobserved due to imperfect detection
\be \begin{split}
\hat{\mathcal{M}}_{01} &= \bra{0_s\,1_\ell}\mathcal{M}\ket{0\,0} = \left( \begin{array}{cc} 0 & 0 \\ \sqrt{(1-\eta)(1-e^{-\gamma\,\Delta t})} & {\color{black} 0} \end{array} \right),
\\ &\approx \sqrt{\Delta t}\,\sqrt{(1-\eta)\,\gamma}\,\hat{\sigma}_- + \mathcal{O}\left( \Delta t^\frac{3}{2} \right).
\end{split} \ee \end{subequations} 
We define $\hat{L} \equiv \sqrt{\gamma}\,\sigma_-$, such that $\hat{\mathcal{M}}_{10} \approx \sqrt{\Delta t}\,\sqrt{\eta}\,\hat{L}$ and $\hat{\mathcal{M}}_{01} \approx \sqrt{\Delta t}\,\sqrt{1-\eta}\,\hat{L}$. 
The operators above represent a complete set of outcomes, in that together they complete a positive operator valued measure (POVM), satisfying
\be \label{POVM_Kraus}
\hat{\mathcal{M}}_{00}^\dag \hat{\mathcal{M}}_{00} + \hat{\mathcal{M}}_{01}^\dag\hat{\mathcal{M}}_{01} + \hat{\mathcal{M}}_{10}^\dag\hat{\mathcal{M}}_{10} = \hat{\mathbb{I}}. 
\ee 
Notice that \eqref{POVM_Kraus} holds both for the general forms of these matrices 
and for the operators approximated to $\mathcal{O}(\Delta t)$. 
Equation \eqref{POVM_Kraus} further implies that $\hat{\mathfrak{Z}} = -\tfrac{1}{2}\hat{L}^\dag\hat{L}$, which is an expected and general property of such jump operators. 

We are now in a position to describe the conditional qubit dynamics (i.e.,~the dynamics of the qubit that can be inferred from the ``click'' or ``no--click'' outcome of the photodetector monitoring the spontaneous emission channel with efficiency $\eta$). 
In the event of a click, we infer a jump of the qubit density matrix $\rho$ to $\ket{g}\bra{g}$, according to
\be 
\rho(t+\Delta t) = \frac{\hat{\mathcal{M}}_{10}\,\rho(t)\,\hat{\mathcal{M}}_{10}^\dag}{\tr{\hat{\mathcal{M}}_{10}\,\rho(t)\,\hat{\mathcal{M}}_{10}^\dag}} = \ket{g}\bra{g}.
\ee
Note that after detecting a jump the final state $\ket{g}$ is obtained irrespective of our estimate of the state prior to the detector click. 
However, we are much more interested in the dynamics that arise when the detector does not click; these dynamics come from the state update
\be \label{eq-no-jump-rho-update}
\rho(t+\Delta t) = \frac{\hat{\mathcal{M}}_{00}\,\rho(t)\,\hat{\mathcal{M}}_{00}^\dag + \hat{\mathcal{M}}_{01}\,\rho(t)\,\hat{\mathcal{M}}_{01}^\dag}{\tr{\hat{\mathcal{M}}_{00}\,\rho(t)\,\hat{\mathcal{M}}_{00}^\dag + \hat{\mathcal{M}}_{01}\,\rho(t)\,\hat{\mathcal{M}}_{01}^\dag}},
\ee
which includes a weighted average over the the two sub-processes that lead to the no-click outcome. 
We may take this expression, and re-write it as a dynamical equation for the density matrix by making expansions to $\mathcal{O}(\Delta t)$. 
In the notation above, we find
\begin{widetext} \be \label{eta-eq} \begin{split}
\dot{\rho} &= \hat{\mathfrak{Z}}\,\rho + \rho\,\hat{\mathfrak{Z}}^\dag + (1-\eta) \hat{L}\,\rho\,\hat{L}^\dag - \rho\,\tr{\hat{\mathfrak{Z}}\,\rho + \rho\,\hat{\mathfrak{Z}}^\dag + (1-\eta) \hat{L}\,\rho\,\hat{L}^\dag} \\
&= \underbrace{(1-\eta)\hat{L}\,\rho\,\hat{L}^\dag  - \tfrac{1}{2}\hat{L}^\dag\hat{L}\,\rho - \tfrac{1}{2}\rho\,\hat{L}^\dag\hat{L}}_\text{Linear (may be written as a Liouvillian)} + \underbrace{\eta\,\rho\,\tr{\hat{L}\,\rho\,\hat{L}^\dag}}_{\text{Non-Linear}},
\end{split} \ee \end{widetext} 
which is appropriate between any detector clicks. 
We have written a trace--preserving equation of motion, which consequently explicitly includes a non-linear term that enforces conservation of probability (i.e.~a normalization). 
In the absence of the non-linear term, the equation can be written in a linear (Liouvillian) form \cite{Minganti_2019, Minganti_2020}. 
The non-linear trace--preserving term arises when $\eta >0$ because the probability of the associated event of no detector click is not equal to one. 
We have identified $\hat{\mathfrak{Z}} = -\tfrac{1}{2}\hat{L}^\dag\hat{L}$ for $\hat{L} = \sqrt{\gamma}\,\hat{\sigma}_-$, and see that the Lindblad master equation is recovered for the unmonitored case (i.e.,~for the unconditional dynamics that arise from $\eta = 0$).

We then arrive at the expressions used in the main text by adding unitary qubit rotations defined by a Rabi drive $\bf{\Omega}$ to the dynamics above (these unitaries function as a control Hamiltonian throughout the main text), resulting in  
$\dot{\rho} = $
\begin{subequations} \be \label{rhodot}
i[\rho,\hat{H}]+(1-\eta)\hat{L}\,\rho\,\hat{L}^\dag  - \tfrac{1}{2}\hat{L}^\dag\hat{L}\,\rho - \tfrac{1}{2}\rho\,\hat{L}^\dag\hat{L} + \eta\,\rho\,\tr{\hat{L}\,\rho\,\hat{L}^\dag}
\ee \be 
\text{with}\quad \hat{H} = \tfrac{1}{2}\left(\Omega_x\,\hat{\sigma}_x + \Omega_y\,\hat{\sigma}_y+\Omega_z\,\hat{\sigma}_z\right).
\ee \end{subequations}
Note that such an expression is intrinsically conceived in the limit of small timesteps. The lack of commutation between the control unitary $\hat{\mathcal{U}} = e^{-i\,\hat{H}\,\Delta t}$ with any of the Kraus operators $\hat{\mathcal{M}}$ may be neglected to $O(\Delta t)$, but will typically contribute non-trivially to $O(\Delta t^\frac{3}{2})$ and beyond. 

With these assumptions in place, the equation of motion can be recast as a dynamical system in the qubit's Bloch coordinates ($\dot{\mathbf{q}} = \tr{\hat{\sigma}_\mathbf{q}\,\rho}$), specifically as
\begin{subequations} \label{conditional-dynamics} \be 
\dot{x} = \mathcal{F}_x(\mathbf{q},\boldsymbol{\Omega}) = \tfrac{1}{2}\gamma\,x\left(\eta(z+1)-1 \right) + z\,\Omega_y-y\,\Omega_z,
\ee \be 
\dot{y} = \mathcal{F}_y(\mathbf{q},\boldsymbol{\Omega}) = \tfrac{1}{2}\gamma\,y\left(\eta(z+1)-1 \right) + x\,\Omega_z-z\,\Omega_x,
\ee \be 
\dot{z} = \mathcal{F}_z(\mathbf{q},\boldsymbol{\Omega}) = \tfrac{1}{2}\gamma\,(1+z)\left(\eta(z+1)-2 \right) + y\,\Omega_x-x\,\Omega_y,
\ee 
for $\mathbf{q} \equiv (x,y,z)^\top$ and $\boldsymbol{\Omega} \equiv (\Omega_x,\Omega_y,\Omega_z)^\top$. 
We may equivalently separate out the parts of the dynamics due to the measurement (leading terms) and the Rabi drive, according to
\be \label{F_factored}
\dot{\mathbf{q}} = \boldsymbol{\mathcal{F}}(\mathbf{q},\boldsymbol{\Omega}) = \gamma\,\tilde{\boldsymbol{\mathcal{F}}}(\mathbf{q}) - \mathbf{q}\times\boldsymbol{\Omega},
\ee
where $\tilde{\boldsymbol{\mathcal{F}}}(\mathbf{q}) =\tfrac{1}{\gamma}\, \boldsymbol{\mathcal{F}}_0$, i.e.,~$\tilde{\boldsymbol{\mathcal{F}}}$ are the un-controlled conditional dynamics, with the decay timescale factored out. 
\end{subequations} 

We may then regroup some terms in \eqref{rhodot} to identify the non--Hermitian Hamiltonian of primary interest. 
Specifically, the dynamics of \eqref{rhodot} may equivalently be expressed as
\begin{subequations} \be \label{NH_ME}
\dot{\rho} = i\,\rho\,\hat{\mathcal{H}}^\dag - i \,\hat{\mathcal{H}}\,\rho+ (1-\eta)\hat{L}\,\rho\,\hat{L}^\dag + \eta\,\rho\,\tr{\hat{L}\,\rho\,\hat{L}^\dag}, 
\ee
where we have defined
\be \label{NHH}
\hat{\mathcal{H}} = \hat{H} - \tfrac{i}{2}\hat{L}^\dag\hat{L} = \frac{1}{2}\left( \begin{array}{cc}
\Omega_z - i\,\gamma & \Omega_x-i\,\Omega_y \\ 
\Omega_x+i\,\Omega_y & -\Omega_z
\end{array} \right),
\ee \end{subequations}
with $\hat{L} = \sqrt{\gamma}\,\hat{\sigma}_-$. 
Eq.~\eqref{NH_ME} is simply a re-writing of \eqref{rhodot} in terms of the NH Hamiltonian. 

\section{Pontryagin Optimal Control: Details and Extended Discussion \label{app-pontryagin}}

\subsection{The Pontryagin Maximum Principle} 

Our use of Pontryagin's principle for optimal control revolves around an action that is generically of the form 
\be 
\mathscr{S} = f(\mathbf{q}_T) + \int_0^T dt\,\left(\mathbf{p}\cdot\dot{\mathbf{q}} - \mathscr{H}(\mathbf{q},\mathbf{p},\mathbf{u}) \right),
\ee
where $\mathbf{q}$ are state coordinates as in the main text, $\mathbf{p}$ are the co-states (i.e.,~conjugate ``momenta'', or Lagrange multipliers, directly analogous to $\boldsymbol{\Lambda}$), and $\mathbf{u}$ are some controller variables. 
Optimization involves solving $\delta \mathscr{S} = 0$, where $\delta \mathscr{S}$
\be \begin{split}
 &= \delta f(\mathbf{q}_T) + \int_0^T\left( \delta[\mathbf{p}\cdot\dot{\mathbf{q}}] - \delta \mathscr{H}(\mathbf{q},\mathbf{p},\mathbf{u}) \right), \\
&= \mathbf{p}\cdot\delta \mathbf{q}|_0^T + \partl{f}{\mathbf{q}_T}{}\delta\mathbf{q}_T + \int_0^T dt \left\lbrace \delta\mathbf{p}\cdot\dot{\mathbf{q}} - \dot{\mathbf{p}}\cdot\delta\mathbf{q} - \delta \mathscr{H} \right\rbrace,
\end{split} \ee 
with
\be
\delta \mathscr{H} = \partl{\mathscr{H}}{\mathbf{q}}{}\delta\mathbf{q} + \partl{\mathscr{H}}{\mathbf{p}}{}\delta\mathbf{p} + \partl{\mathscr{H}}{\mathbf{u}}{}\delta\mathbf{u}. 
\ee
The equations 
\be \label{app-HamEq}
\dot{\mathbf{q}} = \partl{\mathscr{H}}{\mathbf{p}}{},\quad \dot{\mathbf{p}} = -\partl{\mathscr{H}}{\mathbf{q}}{},\quad \partl{\mathscr{H}}{\mathbf{u}}{} = 0
\ee
arise straightforwardly from requiring that the integral contribution goes to zero, independent of the variations $\delta\mathbf{q}$, $\delta\mathbf{p}$, and $\delta\mathbf{u}$. 

This mathematical approach to control, where optimal controls are derived via action extremization, variational calculus or geodesics, has a long history in both classical \cite{LiberzonBook, SchattlerBook} and quantum \cite{DAllessandroBook, Boscain2021} control. 
This includes work on related physical systems, where an action--extremization principle has similarly been used to find extremal--probability measurement records and paths (instead of unitary controllers) for continuously--monitored quantum systems \cite{Chantasri_2013, Weber2014, Chantasri_2015, Jordan_2015Flor, Lewalle_Multipath, Naghiloo_Caustic, Lewalle_Chaos, Chantasri_2021, Karmakar_2022, Lewalle_2020Flor}. 
An application of a similar method towards simultaneous optimization of a diffusive measurement record and unitary controls has recently been carried out \cite{Kokaew_2022}. 

\subsection{Optimal Equations of Motion}

In the main text, we introduce a Hamiltonian for Pontryagin control of our qubit, which reads
\begin{widetext}
\be 
\mathscr{H} = \boldsymbol{\Lambda}\cdot\boldsymbol{\mathcal{F}}(\mathbf{q},\boldsymbol{\Omega}) - \underbrace{\left(\tfrac{\alpha}{2}\left\lbrace (x-x_f)^2 + (y-y_f)^2 + (z-z_f)^2 \right\rbrace + \tfrac{\beta}{2}\left\lbrace \Omega_x^2+\Omega_y^2+\Omega_z^2 \right\rbrace \right)}_{\mathscr{L}(\mathbf{q},\boldsymbol{\Omega})}.
\ee
Note that the actual cost function (i.e.~the part of the Hamiltonian that is not directly implementing the constraint to the dynamical equations $\dot{\mathbf{q}} = \boldsymbol{\mathcal{F}}$) may be recognized as the corresponding Lagrangian $\mathscr{L}$. 
We have chosen a simple form of the cost function in this work; it is possible to follow the same procedure with a modified cost function that may lead to solutions better optimized for a particular task. 

We now go through the optimization calculations based on $\mathscr{H}$ in some detail. 
We consider the drive optimization condition first, finding
\be 
\frac{\partial\mathscr{H}}{\partial\boldsymbol{\Omega}} = 0 \quad\rightarrow\quad \Omega_x^\star = \tfrac{1}{\beta}\left(y\,\Lambda_z-z\,\Lambda_y \right),\quad
\Omega_y^\star = \tfrac{1}{\beta}\left(z\,\Lambda_x-x\,\Lambda_z \right),\quad
\Omega_z^\star = \tfrac{1}{\beta}\left(x\,\Lambda_y-y\,\Lambda_x \right).
\ee
This may remind the reader of expressions for the angular momentum, if the Lagrange multipliers $\boldsymbol{\Lambda}$ are understood as conceptually analogous to linear momenta (i.e.,~the choice of $\mathscr{L}$ above leads to $\boldsymbol{\Omega}^\star = \tfrac{1}{\beta}\,\mathbf{q}\times\boldsymbol{\Lambda}$). 
We emphasize again our remark in the main text that this form of the optimal drive indicates that the choice of $\beta$ is somewhat redundant with the choice we must make in practice about where to bound the Lagranian Manifold, and hence also the Pontryagin energy $\mathscr{E} = \mathscr{H}$.
For this reason we pay relatively little attention below to this ``temperance'' parameter.

Substituting in $\boldsymbol{\Omega}\,\rightarrow\,\boldsymbol{\Omega}^\star = \tfrac{1}{\beta}\,\mathbf{q}\times\boldsymbol{\Lambda}$ leads to
\be \begin{split}
\mathscr{H}^\star &= \boldsymbol{\Lambda}\cdot\left( \boldsymbol{\mathcal{F}}(\mathbf{q},\boldsymbol{\Omega} = 0) - \tfrac{1}{\beta}\, \mathbf{q}\times(\mathbf{q}\times\boldsymbol{\Lambda}) \right) - \tfrac{\alpha}{2}|\mathbf{q}-\mathbf{Q}_f|^2 - \tfrac{1}{2\beta}|\mathbf{q}\times\boldsymbol{\Lambda}|^2 \\
&= \boldsymbol{\Lambda}\cdot\boldsymbol{\mathcal{F}}(\mathbf{q},\boldsymbol{\Omega} = 0) + \tfrac{1}{2\beta}\,\mathbf{q}^2\,\boldsymbol{\Lambda}^2 - \tfrac{1}{2\beta}\,(\mathbf{q}\cdot\mathbf{\Lambda})^2- \tfrac{\alpha}{2}|\mathbf{q}-\mathbf{Q}_f|^2,
\end{split} \ee
which is the Hamiltonian generating optimally--controlled dynamics. 
Note that after substituting in $\boldsymbol{\Omega} = \boldsymbol{\Omega}^\star$ everywhere, some terms from the uncontrolled part of the dynamical equations $\boldsymbol{\mathcal{F}}(\mathbf{q},\boldsymbol{\Omega} = 0) = \boldsymbol{\mathcal{F}}_0$ will remain. 
The Hamiltonian above may equivalently be expanded to read
\be 
\begin{split}
\mathscr{H}^\star =& \tfrac{\gamma}{2}\left\lbrace \left[\eta(z+1)-1\right]\left[x\,\Lambda_x+y\,\Lambda_y \right] + (1+z)\Lambda_z[\eta(z+1)-2] \right\rbrace \\ &~ +\tfrac{1}{2\beta}\left\lbrace \Lambda_x^2(y^2+z^2) + \Lambda_y^2(x^2+z^2) + \Lambda_z^2(x^2+y^2)  \right\rbrace - \tfrac{1}{\beta}\left\lbrace xy\,\Lambda_x\Lambda_y + xz\,\Lambda_x\Lambda_z+yz\,\Lambda_y\Lambda_z \right\rbrace \\ 
& ~- \tfrac{\alpha}{2}\left\lbrace (x-X_f)^2 + (y-Y_f)^2 + (z-Z_f)^2 \right\rbrace.
\end{split} \ee 
The subsequent equations of motion (Hamilton's dynamical equations) now read 
\begin{subequations} \label{eq:dynamics} \be 
\dot{x}^\star = \tfrac{1}{2} \gamma  \,x\lbrace \eta(z+1)-1 \rbrace -\tfrac{1}{\beta}\,x (\Lambda_y y+\Lambda_z z)+\tfrac{1}{\beta}\,\Lambda_x \left(y^2+z^2\right),
\ee \be 
\dot{y}^\star = \tfrac{1}{2} \gamma  \,y \lbrace \eta(z+1)-1\rbrace -\tfrac{1}{\beta}\,y (\Lambda_x x+\Lambda_z z) + \tfrac{1}{\beta}\,\Lambda_y \left(x^2+z^2\right),
\ee \be 
\dot{z}^\star = \tfrac{1}{2} \gamma  (z+1) \lbrace \eta(z+1)-2 \rbrace -\tfrac{1}{\beta}\,z\left( \Lambda_x x + \Lambda_y y \right) + \tfrac{1}{\beta}\,\Lambda_z\left( x^2+y^2 \right) ,
\ee \be 
\dot{\Lambda}_x^\star = \alpha(x-X_f)-\tfrac{1}{2} \gamma\,\Lambda_x \lbrace \eta(z+1)-1 \rbrace  +\tfrac{1}{\beta}\,\Lambda_x (\Lambda_y y+\Lambda_z z) -\tfrac{1}{\beta}\,x \left( \Lambda_y^2+\Lambda_z^2\right),
\ee \be 
\dot{\Lambda}_y^\star = \alpha(y-Y_f)-\tfrac{1}{2} \gamma\,\Lambda_y \lbrace \eta(z+1)-1\rbrace  + \tfrac{1}{\beta}\,\Lambda_y (\Lambda_x x+\Lambda_z z) -\tfrac{1}{\beta}\,y \left(\Lambda_x^2+\Lambda_z^2\right),
\ee \be 
\dot{\Lambda}_z^\star = \alpha(z-Z_f)-\gamma\,  \Lambda_z\lbrace \eta(z+1)-1\rbrace -\tfrac{1}{2} \gamma \, \eta  (\Lambda_y y+\Lambda_x x)+ \tfrac{1}{\beta}\,\Lambda_z(\Lambda_x  x+\Lambda_y  y)-\tfrac{1}{\beta}\,z \left( \Lambda_x^2+\Lambda_y^2\right).
\ee
\end{subequations}
\end{widetext}

It is these equations which are explicitly integrated in the course of finding dynamical solutions. 
Some features of the dynamics are more apparent in this form: For instance, the leading terms on each $\dot{\boldsymbol{\Lambda}}$ equation (i.e.,~$\alpha(\mathbf{q}-\mathbf{Q}_f) - ...$) clearly illustrate how the magnitude and direction of the control drive is adjusted if we find ourselves far from the target state. 
More specifically, $\alpha$ essentially determines how rapidly the control drive can change, such that paths derived with a larger $\alpha$ are capable of greater acceleration.
As shown in Fig.~\ref{fig-pontryagin-accel}, this property is highly useful to reach pure target states under inefficient detection.

In the main text we have worked with a Lagrangian Manifold (LM) of control parameter solutions defined by choosing a volume of initial $\boldsymbol{\Lambda}_i$, corresponding to a volume of initial conditions for optimal control trajectories.  
Solutions in this LM are unique within the full (six--dimensional) phase space of $\mathscr{H}$, but are not necessarily unique when projected into the smaller, physical, $\mathbf{q}$--space.
Regions of the Bloch sphere where multiple solutions meet the given boundary conditions are directly analogous to the caustics that appear in optics and diverse other physical settings, that are described mathematically by catastrophe theory \cite{AlonsoCaustic, BookArnoldClassical, BookArnoldCatastrophe, BookMaslov, Littlejohn1992, Lewalle_Multipath, Naghiloo_Caustic}. 


\subsection{On the Existence of Control Solutions Meeting Particular Boundary Conditions}

We comment here on the simplest case (of pure states with $\eta = 1$), and then on the general mixed--state case with $\eta \leq 1$. 

\subsubsection{Pure State Solutions}

It is easy to see from \eqref{F_factored} that trivial purity--preserving solutions exist for $\eta = 1$ and in the limit $|\boldsymbol{\Omega}| \gg \gamma$. 
If the unitary control dynamics are fast, i.e., they occur on timescales where the decay is negligible, we may factor out the drive magnitude $\dot{\mathbf{q}} = |\boldsymbol{\Omega}|\left(\bar{\Omega}\times\mathbf{q}\right) + \gamma\,\tilde{\boldsymbol{\mathcal{F}}}$ (where $\bar{\Omega}$ is the unit vector setting the rotation axis), and immediately see that
\be
\frac{\dot{\mathbf{q}}}{|\boldsymbol{\Omega}|} = \bar{\Omega}\times\mathbf{q} + \frac{\gamma}{|\boldsymbol{\Omega}|} \tilde{\boldsymbol{\mathcal{F}}} \approx \bar{\Omega}\times\mathbf{q}.
\ee
In other words, in the limit of fast drive ($\gamma \ll |\boldsymbol{\Omega}|$), the decay dynamics can be treated as a perturbation, and the trivial linear control problem arising from Rabi drive alone is recovered in the limit where the controls are applied very strongly.  
We consequently conclude that our ability to map arbitrary pure states to other arbitrary pure states in the ideal ($\eta = 1$) case is limited only by the controller power. 

The mixed state case is less straightforward. We will find that it is useful to supplement the Cartesian representation of our dynamics above with a spherical coordinate representation before considering the mixed state solutions in detail. 

\subsubsection{Dynamics and Control in Spherical Coordinates}

Let us convert \eqref{conditional-dynamics} to spherical coordinates $x = R\,\cos\phi\,\sin\theta$, $y = R\,\sin\phi\,\sin\theta$, and $z = R\,\cos\theta$, thereby representing the dynamics by 
\begin{subequations} \label{Fsphere} \be \label{Rdot} \begin{split}
\dot{R} = \gamma\big\lbrace &  \tfrac{1}{2} \cos \theta \left(\eta(1+R^2)-2\right) \\ \quad\quad\quad & +\tfrac{1}{4} R (\eta -1)(\cos (2 \theta )+3) \big\rbrace,
\end{split} \ee \\ \be \begin{split}
\dot{\theta} = &-\frac{\gamma  (\eta +(\eta -1) R \cos \theta -2) \sin \theta }{2 R}\\ &-\Omega_x \sin \phi +\Omega_y \cos\phi,
\end{split} \ee \be
\dot{\phi} = \Omega_z - (\Omega_x \cos \phi +\Omega_y \sin \phi )\cot \theta.
\ee \end{subequations} 
A large part of our interest in the spherical form of the problem stems from the dynamics of state purity $\mathcal{P} = \tr{\rho^2}$.
This expression is closely related to the radial equation of motion, i.e., $\dot{\mathcal{P}} = 2\,\tr{\rho\,\dot{\rho}} = R\,\dot{R}$.
This can equivalently be written as $\dot{\mathcal{P}} = $
\be 
\tfrac{1}{2}\gamma\,  \left[(x^2+y^2)\lbrace \eta(1+z)-1 \rbrace +z (z+1) \lbrace \eta(1+z)-2 \rbrace \right]. 
\ee
Since these dynamics related to changes in state purity are independent of $\phi$, they can be completely represented within a cross-sectional plane of the Bloch sphere; graphical representations appear in Fig.~\ref{fig-PurDot}.

The control problem as a whole can also be recast in spherical coordinates, where the change from Cartesian $\mathbf{q} = (x,y,z)$ and $\boldsymbol{\Lambda} = (\Lambda_x,\Lambda_y,\Lambda_z)$ to spherical $\mathbf{q} = (R,\theta,\phi)$ and $\boldsymbol{\Lambda} = (\Lambda_R,\Lambda_\theta,\Lambda_\phi)$ is performed via canonical transformation:
\begin{widetext} \be 
\begin{array}{ccc}
x & \rightarrow & R\,\cos\phi\,\sin\theta ,\\
y & \rightarrow & R\,\sin\phi\,\sin\theta ,\\
z & \rightarrow & R\,\cos\theta ,\\
\Lambda_x & \rightarrow & \Lambda_R \,\sin\theta\,\cos\phi + (\Lambda_\theta/R)\cos\theta\,\cos\phi - (\Lambda_\phi/R)\csc\theta\,\sin\phi ,\\
\Lambda_y & \rightarrow & \Lambda_R\,\sin\theta\,\sin\phi + (\Lambda_\theta/R)\cos\theta\,\sin\phi + (\Lambda_\phi/R)\csc\theta\,\cos\phi ,\\
\Lambda_z & \rightarrow & \Lambda_R\,\cos\theta - (\Lambda_\theta/R)\sin\theta . 
\end{array}
\ee \end{widetext} 
Applying the above transformation to the Cartesian form of $\mathscr{H}$ preserves the Poisson bracket between all dynamical variables, and is equivalent to re-assembling the Hamiltonian in spherical coordinates as follows
\be 
\mathscr{H} = \Lambda_R\,\dot{R} + \Lambda_\theta\,\dot{\theta} + \Lambda_\phi\,\dot{\phi} - \mathscr{L}(\mathbf{q},\boldsymbol{\Omega}).
\ee
Note that the equations for the optimal control (i.e.,~from solving $\partial_{\boldsymbol{\Omega}}\mathscr{H} = 0$) now read
\begin{subequations} \be 
\Omega_x^\star = - \Lambda_\theta\,\sin\phi -\Lambda_\phi \,\cot\theta\, \cos\phi ,
\ee \be 
\Omega_y^\star = \Lambda_\theta\,\cos\phi-\Lambda_\phi\,\cot\theta\,\sin\phi ,
\ee \be 
\Omega_z^\star = \Lambda_\phi .
\ee \end{subequations}
These are quite helpful in constructing simulations, in that they formalize an intuition that the co-state variable $\Lambda_R$ plays no role in the controller trajectory (because $\Lambda_R$ corresponds to changes in purity, which the unitary controller cannot implement directly). 
Consequently, we may work with a two--dimensional Lagrangian Manifold (LM), completely defined by an initial mesh of $\Lambda_\theta$ and $\Lambda_\phi$, rather than a three--dimensional one in $\Lambda_x$, $\Lambda_y$, and $\Lambda_z$. This reduction in the dimensionality of the integrated manifold considerably simplifies the numerics. 

\subsubsection{Mixed State Solutions} 

\begin{figure*}
    \centering
\includegraphics[width = \textwidth]{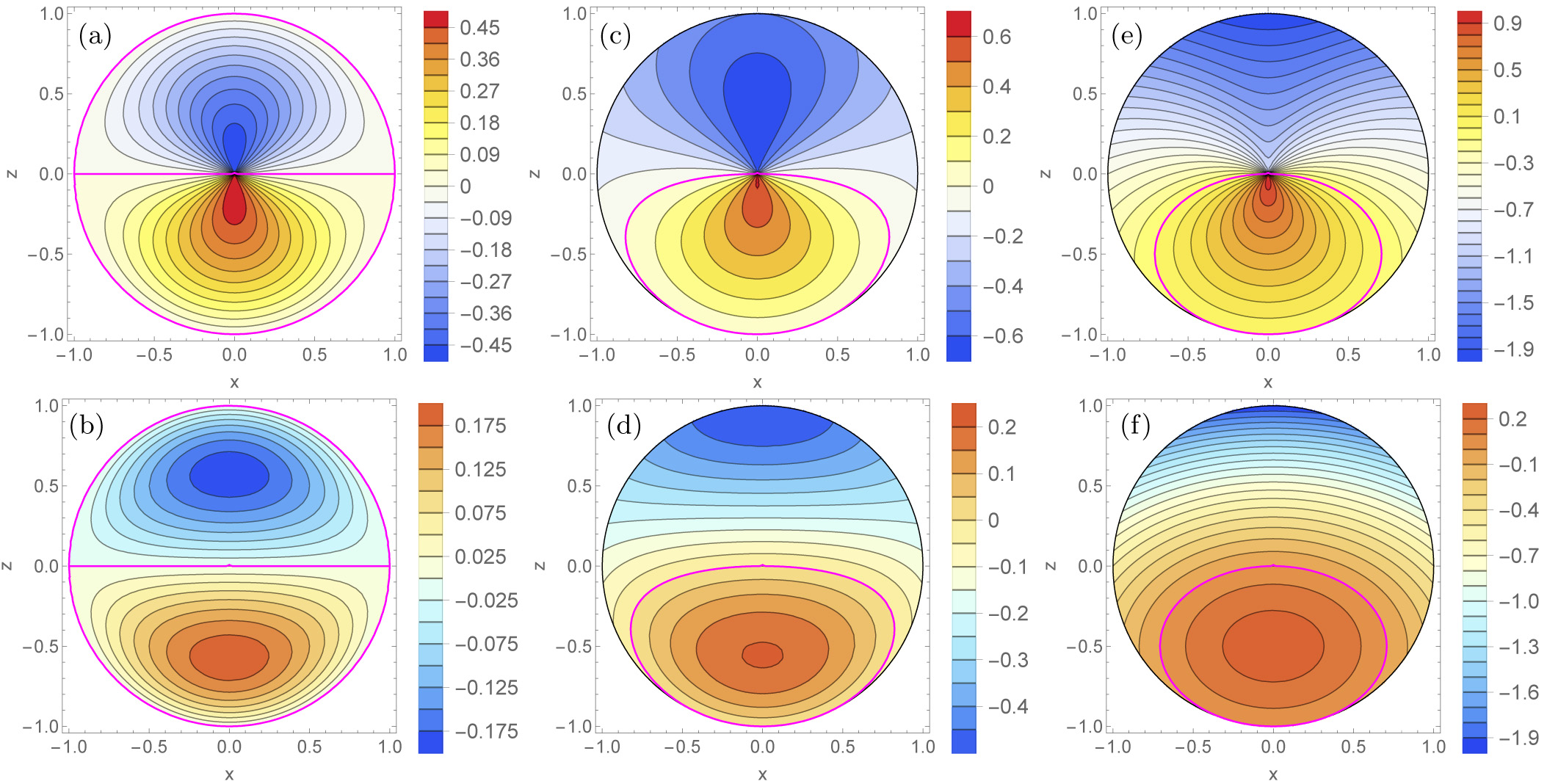}
    \caption{
    Graphical representation of the change in purity for mixed state no--jump dynamics \eqref{conditional-dynamics}. We plot the rate of change of the Bloch vector length, $\dot{R}$ \eqref{Rdot}, on the top (panels a,c,e) and the rate of change of the purity $\dot{\mathcal{P}}$ on the bottom (panels b, d, f). 
    Results are shown for detection efficiencies $\eta = 1$ (panels a, b), $\eta = 0.75$ (panels c, d), and $\eta = 0$ (panels e, f). 
    Contour plots show lines of constant $\dot{R}(x,z)$ or $\dot{\mathcal{P}}(x,z)$, such that the dynamical flow will run perpendicular to the contour lines, i.e.~along the gradient. 
    Both $\dot{R}$ and $\dot{\mathcal{P}}$ values are given in units of $\gamma$. 
    The $\dot{R} = 0$ and $\dot{\mathcal{P}} = 0$ lines are shown in magenta; purity increases within the magenta boundary, and decreases outside of it.
    We can understand that purity decreases in the upper half of the Bloch sphere (i.e.,~closer to $\ket{e}$) in the inefficient conditional dynamics, and does so faster for worse efficiencies. 
    On the other hand, the dynamical attraction towards $\ket{g}$ can be used to increase purity for states in a large region between the equator and $\ket{g}$, across a wide range of measurement efficiencies. 
    These results generically imply that control protocols can reach states with $\mathcal{P}_f > \mathcal{P}_i$ if the target trajectory spends sufficient time traversing the region where $\dot{\mathcal{P}} > 0$. 
    }
    \label{fig-PurDot}
\end{figure*}

The rate of change of the state purity $\dot{\mathcal{P}}$ is especially important to us in the cases of inefficient detection, $\eta < 1$, as well as an initially impure qubit state, because the amount of evolution time spent in regions of $\dot{\mathcal{P}} > 0$ will upper-bound the final state purity (and thereby constrain which final states are attainable). 
Recall that unitary dynamics (i.e.,~$\boldsymbol{\Omega}$) cannot directly change the state purity.
Purity is here increased (decreased) purely through the acquisition (loss) of information the qubit has shared with its optical environment via spontaneous emission. 
What we can do, however, is to devise controller trajectories that drive the system to a state where the no-click (no-jump) measurement information leads to purity increase, i.e.,~a measurement induced change in purity that depends on the qubit state. In this manner we can use our controls to indirectly affect $\mathcal{P}$. 

Qualitative aspects of the relevant dynamics may be immediately inferred from Fig.~\ref{fig-PurDot}. 
For instance, the line $\dot{\mathcal{P}} = 0$ necessarily goes through $\ket{g}$ irrespective of $\eta$, and this point is the only state where $\dot{\mathcal{P}} = 0$ \emph{and} $\mathcal{P} = 1$ for $\eta < 1$. 
It follows that the jumpless dynamics (including the Lindbladian dynamics without detection) can only asymptotically approach $\mathcal{P} = 1$. 
More specifically, the dynamics without drive ($\boldsymbol{\Omega} = 0$) decay as
\be 
z(t) = \frac{2\,u_0}{\eta\,u_0 - e^{\gamma\,t}(\eta\,u_0-2)} - 1
\ee
for $u_0 = 1+z_0 \in [0,2]$.
These solutions asymptotically approach $z = -1$ in the long--time limit for all parameters except for $z_0 = 1$ (initial $\ket{e}$) and $\eta=1$ \cite{Lewalle_2020Flor, Tan_2017}. 
Therefore, if one needs to significantly increase state purity, there are two options to accomplish this (which are not mutually incompatible).
The first is to drive the trajectory close to $\ket{g}$ to the greatest extent possible, then wait long enough for the jumpless dynamics to purify the state to the desired degree, and then quickly drive the qubit from near $\ket{g}$ to the target state. 
The second is to alternatively actually measure a click, which resets the qubit directly to $\ket{g}$, at which point the attainable state purity is limited solely by the speed of the controls. Note that a slower trajectory from $\ket{g} \rightarrow \mathbf{Q}_f$ will have more time to lose purity, and therefore have a more constrained range of accessible final states (recall Fig.~\ref{fig-Pontryagin-mixed}). 

We have demonstrated in Fig.~\ref{fig-pontryagin-accel} that our optimal equations of motion are, especially for sufficiently large $\alpha$, capable of generating the type of solutions we have just described. 
A larger value of the ``agility" parameter $\alpha$ helps with this, because it allows our controller to ``accelerate'' towards the target state a short time before we intend to reach it, and to stay near $\ket{g}$ to retain purity prior to that. 
Additional figures and animations illustrating this effect can be found in the Supplementary Materials, described in Appendix \ref{app-SM-list}.  

\section{Supplementary Plots and Animations \label{app-SM-list}}

We provide supplementary animations in the accompanying slide deck \cite{Sup_Mat_Ani_Revised}.
These animations are designed to offer clear visual illustrations of the following points: 
\begin{enumerate}
    \item (Slide 3)  We illustrate bi--orthogonal states on the Bloch sphere (see Ashida et al., 2020), in preparation for subsequent animations.
    \item (Slides 5--6) We demonstrate how adiabatic following can succeed or fail, using closed loop parameter trajectories of duration $T = 100\,T_1$. Adiabatic following works just as in the Hermitian case if our entire trajectory stays on the stable eigensheet (slide 5), but fails when we encircle an EP and thereby force a stability exchange (slide 6) (see \cite{Nenciu_1992}). 
    \item (Slides 8--9) Stability exchange when encircling an EP can also be understood as resulting in chiral state exchange. We illustrate a situation where we perform a clockwise encircling that stays on the stable eigensheet (slide 8), in contrast with a counter--clockwise encircling that attempts to follow the lossy eigensheet \cite{Zhong2018}. 
    \item (Slides 11--12)  Pure state optimal manifold dynamics from Pontryagin control cover the entire surface of the Bloch sphere over moderate time intervals for moderate manifold energy boundaries (which limit the maximum Rabi rotation rates $|\boldsymbol{\Omega}|$ available to the controller). These are animated versions of each panel in Fig.~\ref{fig_pure_control}.
    \item {\color{black} (Slides 14--18)} Pontryagin control under inefficient measurement ($\eta <1$) leads to difficulties in the exact targeting of pure final states. 
We show additional plots and animations detailing this decoherence. We then develop further examples illustrating here that the issue can be mitigated, demonstrating attainment of increasingly pure final states using increasing controller agility. Larger values of $\alpha$ can significantly improve the purity of final states reached by trajectories $t \gtrsim T_1$; our animations here supplement Figs.~\ref{fig-Pontryagin-mixed} and \ref{fig-pontryagin-accel}. 
\end{enumerate}

\bibliography{refs}
\end{document}